\documentclass[%
reprint,
aps,
prb,
superscriptaddress,
]{revtex4-1}
\usepackage{enumerate}

\usepackage[breaklinks=true]{hyperref}
\hypersetup{colorlinks=true, linkcolor=blue, urlcolor=blue, citecolor=blue}
\usepackage{graphicx}
\usepackage{amsmath} \usepackage{amssymb}
\usepackage{accents} \usepackage{mathrsfs} \usepackage{setspace}
\usepackage{color}
\definecolor{dgreen}{RGB}{00, 120, 00} \definecolor{dblue}{RGB}{00, 00, 220}
\definecolor{lgreen}{RGB}{46, 139, 87} 
\usepackage{ulem}

\newcommand{\bs}[1]{\boldsymbol{#1}} \newcommand{\ul}[1]{\underline{#1}}
\newcommand{\tr}[1]{\mathrm{Tr}\left[#1\right]}

\newcommand{\ut}[1]{\undertilde{#1}} 
\newcommand{\ve}[0]{\varepsilon} \newcommand{\vphi}[0]{\varphi}

\begin{document}

%
%

\title{Fulde-Ferrell-Larkin-Ovchinnikov state in a superconducting thin film 
attached to a ferromagnetic cluster}

\author{Shu-Ichiro Suzuki}
\affiliation{MESA\hspace{-1mm}+ Institute for Nanotechnology, University of Twente, 
7500 AE Enschede, The Netherlands}

\author{Takumi Sato}
\affiliation{Department of Applied Physics, Hokkaido University, 
Sapporo 060-8628, Japan}

\author{Alexander A. Golubov}
\affiliation{MESA\hspace{-1mm}+ Institute for Nanotechnology, University of Twente, 
7500 AE Enschede, The Netherlands}

\author{Yasuhiro Asano}
\affiliation{Department of Applied Physics, Hokkaido University, 
Sapporo 060-8628, Japan}

\date{\today}

\begin{abstract}
We study theoretically the Fulde-Ferrell-Larkin-Ovchinnikov (FFLO)
states appearing locally in a superconducting thin film with a small
circular magnetic cluster. 
By solving the Eilenberger equation in two dimensions, we calculate the pair potential, pairing correlations, free-energy density, and quasiparticle density of states for various cluster sizes and exchange potentials. Increasing the exchange potential and cluster size leads to a higher number of nodes in the pair potential. 
Although the free-energy density beneath the ferromagnet locally exceeds the normal-state value, the FFLO states are stabilized by the superconducting condensate away from the magnetic cluster. 
Analyzing the pairing-correlation functions, we show that the spatial variation of the spin-singlet $s$-wave pair potential generates $p$-wave Cooper pairs. These odd-frequency Cooper pairs play a dominant role in governing the inhomogeneous subgap spectra observed in the local density of states. Furthermore, we propose an experimental method for the detection of local FFLO states by analyzing the quasiparticle density of states.
\end{abstract}

\pacs{pacs}

\thispagestyle{empty}

\maketitle

\section{Introduction}

{
In the presence of a Zeeman field, a spin-singlet Cooper pair has 
center-of-mass momentum, causing the oscillation in the superconducting 
pair potential in real space. 
These oscillating superconducting states, known as
Fulde-Ferrell\cite{fulde_fflo} (FF) and
Larkin-Ovchinnikov\cite{larkin_fflo} (LO) states, were proposed in
the 1960s but remained elusive until recently. 
Several experiments have indicated the
possibility of observing the FFLO state in
layered organic superconductors, 
\cite{
PhysRevLett.99.187002,
PhysRevLett.100.117002,
PhysRevB.85.214514,
PhysRevB.86.064507,
croitoru2013fulde,
mayaffre2014evidence,
PhysRevLett.116.067003,
PhysRevB.103.L220501}
FeSe,\cite{kasahara:prl2020, kasahara:prl2021} 
and other materials.\cite{
kinjo2022superconducting,
PhysRevLett.121.157004}
Theoretical studies~\cite{burkhart:annphys1994,matsuo:jpsj1998} have
suggested that the oscillation of the pair potential makes the
superconducting state unstable. On the other hand, the oscillation of
\textsl{pairing correlations} has been extensively discussed in
superconductor/ferromagnet (SF)
junctions.\cite{bulaevskii:jetplett1977, buzdin:jetplett1982,
ryazanov:prl2001, kontos:prl2002, mironov:prl2012}
Although the pair potential is absent in a ferromagnet, the proximity effect
allows non-zero pairing correlations. The success of these
studies highlights SF junctions as a suitable system for investigating
the nature of FFLO states. 
}
{
For instance, in recent experiments,\cite{cren:naturecommun2017, 
menard2019yu,
menard2019isolated}
a superconducting state with a
significant exchange potential was achieved by creating a hybrid
structure consisting of a circular ferromagnetic cluster attached to a
thin superconducting film. 
This configuration suggests the possibility of the formation of a
localized FFLO state within the superconducting region beneath the
ferromagnetic cluster.
}

The effects of magnetic objects embedded in a superconductor on
superconducting states have been studied since the 1960s.  It is
well-established that magnetic impurities decrease the
superconducting transition temperature
$T_c$,\cite{abrikosov:jetp1961} form an impurity band below the
superconducting gap,\cite{yu:actphys1965, shiba:ptp1968,
rusinov:jetplett1969, yazdani:science1997} and are an element for
realizing the topologically nontrivial superconducting
state.\cite{choy:prb2011,Nadjerge:science2014} {These effects depend
not only on the impurity concentration and the amplitude of the
magnetic moments. The size of a magnetic object is also an important
factor.}
{
A point-like magnetic impurity suppresses the pair potentials locally
but the suppression in not significant.\cite{rusinov:jetp1969, fominov:prb2016} 
An finite-size magnetic cluster, on the other hand, 
changes the sign of the pair potential.\cite{salkola:prb1997, flatte:prl1997, balatsky:rmp2006, shu:prb2022}
}
In our previous paper,\cite{shu:prb2022} we have shown that odd-frequency
Cooper pairs surrounding the magnetic cluster are responsible for the
sign change of the pair
potential.\cite{kuzmanovski:prb2020,perrin:prl2020} 
It is widely
accepted that odd-frequency Cooper pairs exist
locally in various SF hybrid
structures \cite{bergeret:prl2001} and play an essential role in
various physical phenomena. For instance, the Josephson current
through a half-metallic ferromagnet\cite{keizer:nature2006,
asano:prl2007sfs, asano:prb2007sfs, braude:prl2007,
robinson:science2010, birge:prl2010, anwar:prb2010} is attributed to
odd-frequency pairs {induced} in the ferromagnet. 
However, the effects of odd-frequency Cooper pairs on the FFLO state are currently not well understood.  We address this issue in the present paper.

\begin{figure}[b]
	\includegraphics[width=0.46\textwidth]{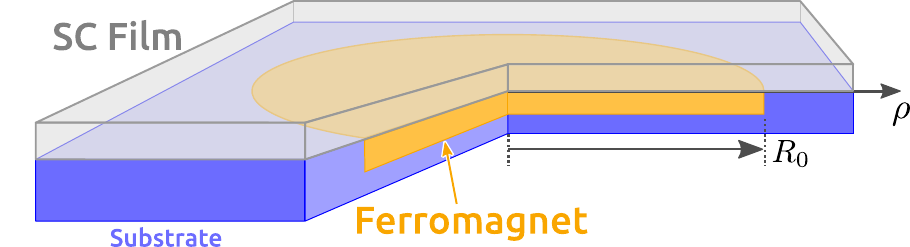}
	\caption{Schematic picture of the system. A thin film of an $s$-wave superconductor is deposited on a
	circular-shaped magnetic cluster. 
	The radius of the magnetic cluster $R_0$ is of the order of the coherence length $\xi_0$. }
  \label{fig:sche}
\end{figure}

{
We study the characteristic features of the local FFLO states in a
superconducting thin film to which a circular magnetic cluster is
attached (see Fig.~\ref{fig:sche}).  
Using the Eilenberger theory, we 
}
calculate the pair potential, the pairing correlation functions, the
free-energy density, and the quasiparticle local density of states (LDOS) for
several choices of the cluster size and exchange potential. 
The results
indicate that the superconducting condensate away from the magnetic
segment stabilizes the local FFLO states.  We also conclude that
odd-frequency Cooper pairs support the sign change of the pair
potential in real space and govern the inhomogeneous subgap spectra of
the LDOS.

This paper is organized as follows. In Sec.~II, we explain {an SF}
structure considered in this paper and theoretical tools to analyze the superconducting {state}.
We {discuss} the numerical results in two-dimension in Sec.~III. 
We also discuss numerical results in one dimension in Sec.~IV.
The {conclusions  are} given in Sec.~V.

\section{Model and formulation}\label{sec:model}

We consider the hybrid structure shown in Fig.~\ref{fig:sche}.
A circular magnetic cluster is attached to an infinitely large superconducting thin film in the $x$-$y$ plane.
The effects of the magnetic cluster are considered through the exchange
potential proximities into the superconducting film,
\begin{align}
	\bs{V}(\bs{r}) = V_0 {\Theta(R_0-\rho)} \bs{e}_z, 
  \label{eq:mag_def}
\end{align}
where $\rho=\sqrt{x^2+y^2}$ and $R_0$ is the radius of the magnetic cluster
and $\Theta(\rho)$ is the Heaviside step function.
Therefore, $\rho=R_0$ indicates the boundary between the magnetic segment 
and the nonmagnetic segment on the superconducting thin film. 

We examine the properties of the superconducting states 
utilizing the quasiclassical Eilenberger theory.\cite{eilenberger:zphys1968} 
The Green's functions obey the Eilenberger equation: 
\begin{align}
  & i \bs{v}_F \cdot \bs{\nabla} \check{g}
	+ \left[ \, i\omega_n \check{\tau}_3+\check{H},~\check{g} \right]_-
	= 0, 
	\label{eq:Eilen}
	\\[2mm]
  & 	\check {g} = \left( \begin{array}{rr}
	 \hat{    g}  &  
	 \hat{    f}  \\[1mm]
	-\hat{\ut{f}} & 
	-\hat{\ut{g}} \\
	\end{array} \right), 
	\hspace{4mm}
	 \check {H} 
	 = \left( \begin{array}{cc}
	 \bs{V} \cdot \hat{ \bs{\sigma}}&  
	 \hat{    \Delta } \\[1mm]
	 \hat{\ut{\Delta}} & 
	 \bs{V} \cdot \hat{ \bs{\sigma}}^* \\
	\end{array} \right), 
\end{align}
where 
$\check{g}= \check{g} (\bs{r},\bs{k},i\omega_n)$ is the quasiclassical
Green's function in the Matsubara representation, 
$\hat {\Delta}(\bs{r})$ is the pair potential, 
$\bs{v}_F = v_F \bs{k}$ is the Fermi velocity, 
and we assume the junction is in equilibrium. 
The \textit{undertilde} function
\begin{align}
  \hat{\undertilde{K}}
  (\bs{r},\bs{k},i \omega_n) \equiv \hat{K}^\ast (\bs{r},-\bs{k},i \omega_n),
\end{align}
represents particle-hole conjugation of $\hat{K}
(\bs{r},\bs{k},i \omega_n)$, 
where the unit vector $\bs{k}$ points the direction of the Fermi momentum. 
In this paper, the accents {$\check{\cdot}$ and
$\hat{\cdot}$ mean}
matrices in particle-hole and those in spin space, respectively.  
The Pauli matrices in these spaces are denoted
by $\check{\tau}_j$ and $\hat{\sigma}_j$ with $j \in \{1, 2, 3 \}$. 
The identity matrices are represented by
$\check{\tau}_0$ and $\hat{\sigma}_0$.  
Throughout this paper, we use the system of units $\hbar=k_B=c=1$,
where $k_B$ is the Boltzmann constant and $c$ is the speed of light.

The Eilenberger equation \eqref{eq:Eilen} can be simplified by the 
Riccati parameterization.\cite{schopohl:arxiv1998,
Schopohl_PRB_95, Eschrig_PRB_00, Eschrig_PRB_09} The Green's function
can be
expressed in terms of the coherence function $\hat{\gamma}=
\hat{\gamma}(\bs{r}, \bs{k}, i \omega_n)$:
\begin{align}
	& \check {g} = 2 \left( \begin{array}{rr}
	 \hat{\mathcal{G}} &  
	 \hat{\mathcal{F}} \\[1mm]
	-\hat{\ut{\mathcal{F}}} & 
	-\hat{\ut{\mathcal{G}}} \\
	\end{array} \right) - \check{\tau}_3, 
	\\
	& \hat{\mathcal{G}} = (1-\hat{\gamma} \hat{\ut{\gamma}})^{-1},\hspace{6mm}
	\hat{\mathcal{F}} = (1-\hat{\gamma} \hat{\ut{\gamma}})^{-1}\hat{\gamma}.
\label{eq:Ric-Para}
\end{align}
The equation for $\hat{\gamma}$ is reduced into the Riccati-type
differential equation: 
\begin{align}
  (i \bs{v}_F \cdot \bs{\nabla} + 2 i \omega_n ) \hat{\gamma}
	+ (\bs{V} \cdot \hat{\bs{\sigma}}) \hat{\gamma} 
	- \hat{\gamma} (\bs{V} \cdot \hat{\bs{\sigma}}^*) &
	\notag \\
	-\hat{\Delta}
	+ \hat{\gamma} \hat{\ut{\Delta}} \hat{\gamma}= 0& . 
	\label{eq:Riccati03}
\end{align}
For a spin-singlet $s$-wave superconductor ($\hat{\Delta} = \Delta i\hat{\sigma}_2$)
under the exchange potential in Eq.~(\ref{eq:mag_def}),
the anomalous Green's function and the coherence function are 
are represented by 
\begin{align}
	&     \hat{f}  
	=    i  (f_0 + f_3 \hat{\sigma}_3) \hat{\sigma}_2,
	\hspace{6mm}
		\ut{\hat{f}} = -i \hat{\sigma}_2 
		( \ul{f}_0 + \ul{f}_3 \hat{\sigma}_3 ), \\
	&     \hat{\gamma}  
	=    i  (\gamma _0 + \gamma_3 \hat{\sigma}_3) \hat{\sigma}_2,
	\hspace{6mm}
		\ut{\hat{\gamma}} = -i \hat{\sigma}_2 
		( \ul{\gamma}_0 + \ul{\gamma}_3 \hat{\sigma}_3 ), 
\end{align}
where 
$\ul{f}_0(\bs{k}) = -f_0^*(-\bs{k})$ and 
$\ul{f}_3(\bs{k}) =  f_3^*(-\bs{k})$. 
Equation~\eqref{eq:Riccati03} can be reduced to 
\begin{align}
  & \bs{v}_F \cdot \bs{\nabla}  \gamma_0
	+ 2  (\omega_n \gamma_0-i V \gamma_3)  
	\notag \\[-1mm] &\hspace{31.5mm}
		- \Delta 
	  + \Delta^* [\gamma_0^2 + \gamma_3^2 ]
		= 0,
	\label{eq:Riccati07}
	\\
  & \bs{v}_F \cdot \bs{\nabla} \gamma_3
	+ 2  (\omega_n \gamma_3-i \gamma_0 V)
	  + \Delta^* 
	[2\gamma_0 \gamma_3 ]= 0, 
\end{align}
The coherence functions far from the magnetic cluster (i.e., $\rho \gg R_0$), 
$\bar{\gamma}(\bs{k}, i \omega_n) $ is calculated as
\begin{align}
  \bar{\gamma}_0 = &
	\frac{{\bar{\Delta}} }
	{\omega_n + \sqrt{\omega_n^2 + |{\bar{\Delta}}|^2} }, 
	\hspace{6mm}
  \bar{\gamma}_3 = 0, 
\end{align}
where $\bar{\cdots}$ means 
the value in the homogeneous region. 
In this paper, we assume the homogeneous superconductivity at $\rho \gg
R_0$.\footnote{In our configuration, a vortex state could be a possible solution of the
Eilenberger equation. However, the vortex has typically a higher energy than the
homogeneous state. Therefore, in this paper, we focus on the
homogeneous superconductivity at $\rho \gg R_0$.}

The spatial profile of the pair potential is determined by the
solving the gap equation self-consistently
\begin{align}
	& \Delta(\bs{r})
	=
	2 \lambda N_0 \frac{\pi}{i \beta} \sum_{\omega_n}^{\omega_c}
	\langle f_0(\bs{r},\bs{k}',i \omega_n) \rangle, 
	\label{eq:gap}
	\\
	& \lambda 
	= \frac{1}{2 N_0}
	\left[
	\ln\frac{T}{T_c} 
	+ \sum_{n=0}^{n_c}
	\frac{1}{n+1/2}
	\right]^{-1}, 
\end{align}
where $\beta=1/T$, $T_c$ is the critical temperature, $N_0$ is the
density of the states (DOS) in the normal state at the Fermi energy,
$\omega_c$ is the high-energy cut-off, and $n_c=[\omega_c/2\pi T_c]$.
The angle average on the Fermi surface is denoted by 
$\langle \cdots \rangle = \int_{-\pi}^{\pi} \cdots d \vphi_k /2\pi$,
where $k_x = \cos \vphi_k$ and $k_y = \sin \vphi_k$ with $\vphi_k$ being the
azimuthal angle in the momentum space. 
The coordinate in real space is parameterized as 
$\bs{r}=(\rho \cos\phi_r, \rho \sin\phi_r)$. 
The LDOS can be calculated from the diagonal parts of the Green's
function, 
\begin{align}
	& N(\bs{r},\ve)
	=
	N_0 
	\langle \tr{ \hat{g} (\bs{r},\bs{k}',i \omega_n) } 
	\rangle|_{i \omega_n \to \ve + i\delta}, 
	\label{eq:LDOS}
\end{align}
where $\delta$ is the smearing factor.

\begin{figure*}[t]
	\includegraphics[width=0.98\textwidth]{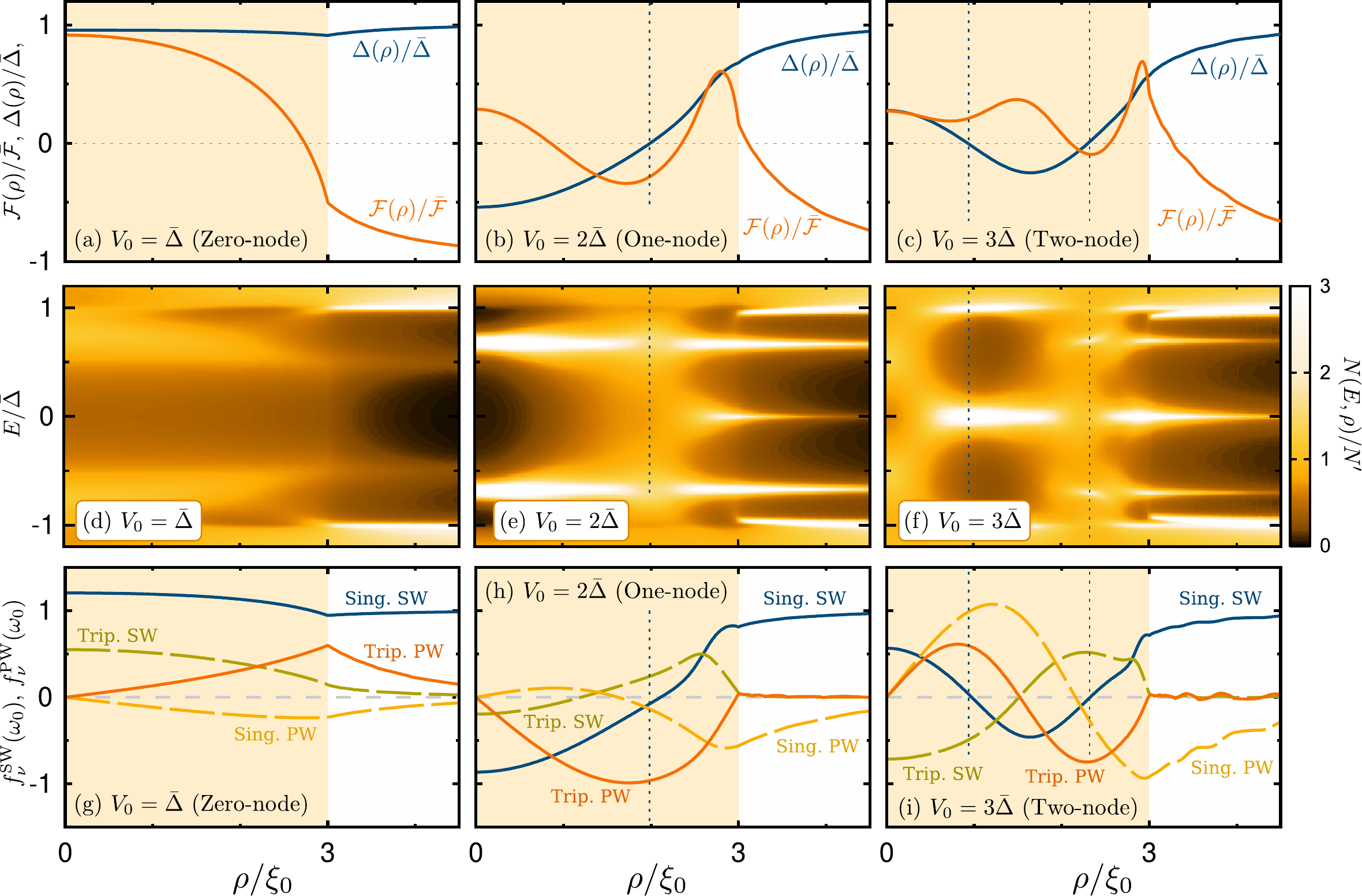}
	\caption{ 
	Spatial profiles of the pair potential
		$\Delta(\rho)$ and free-energy density $\mathcal{F}(\rho)$ 
		for (a) $V_0=\bar{\Delta}$, (b) $2\bar{\Delta}$, and (c) $3\bar{\Delta}$. 
		The results are normalized to their bulk values: 
		$\bar{\Delta}$ and $\bar{\mathcal{F}}$. 
		The positive free-energy density means that the normal state is
		stabler than a superconducting state locally. 
		The vertical dotted lines indicate the places 
		of nodes in the pair potential.
		The shaded areas indicate the area beneath the magnetic cluster with $R_0 = 3\xi_0$.
		Local densities of states (LDOS) for 
 (d) $V_0=\bar{\Delta}$, (e) $2\bar{\Delta}$, and (f) $3\bar{\Delta}$. 
 The LDOS are normalised to $N'=2N_0$. 
		Pairing correlation functions for 
    (g) $V_0=\bar{\Delta}$, (h) $2\bar{\Delta}$, and (i) $3\bar{\Delta}$. 
		The solid (broken) lines indicate the results of 
		even-frequency (odd-frequency) pairing correlations.
		The temperature and the cutoff energy are set to $T=0.2T_c$ and	$\omega_c=6 \pi T_c$. 
		}
  \label{fig:del_Vdep}
\end{figure*}

The pairing correlation function is decomposed into four dominant components.
The spin-singlet $s$-wave component 
\begin{align}
	& f_{0}^{\mathrm{SW}}(\rho, i\omega)
	= \frac{1}{2}
	\langle 
	\mathrm{Tr}[ {(i \hat{\sigma}_2)^\dagger} 
	\hat{f}(\rho, \bs{k}, i \omega_n) 
	] 
	\rangle, 
\end{align}
is the most dominant far from the magnetic cluster and is linked to the pair potential
as shown in Eq.~(\ref{eq:gap}).
The spin-triplet $s$-wave component 
\begin{align}
	& f_{3}^{\mathrm{SW}}(\rho, i\omega)
	= \frac{1}{2}\langle 
	\mathrm{Tr}[ {(i \hat{\sigma}_3 \hat{\sigma}_2)^\dagger} 
	\hat{f}(\rho, \bs{k}, i \omega_n)
  ] \rangle, 
\end{align}
is generated by the exchange potential and belongs to odd-frequency symmetry class.
In what follows, we display the calculated results at 
$\phi_r=0$ because these components are isotropic in real space and independent of $\phi_r$. 
In addition to $s$-wave components, 
odd-parity $p$-wave components also appear in the FFLO state
since the spatial variation of the pair potential breaks inversion
symmetry locally.\footnote{
{
Inversion symmetry is broken
also by the spatially oscillating pair potential in the FFLO states.
When the inversion symmetry is broken, 
parity is no longer a well-defined
symmetry index. In other words, the parity mixing among even- and
odd-parity pairing functions is allowed. 
Therefore, odd-parity pairs exist in the FFLO state of a spin-singlet
even-parity superconductor.}
}
Along the $x$ direction $\boldsymbol{r}=(\rho, 0)$, 
two $p_x$-wave components are generated as a result of breaking 
inversion symmetry in the $x$ direction:
spin-singlet $p_x$-wave component $f_0^{\mathrm{PW}}$ and 
spin-triplet $p_x$-wave component $f_3^{\mathrm{PW}}$ defined by
\begin{align}
	& f_{\nu}^{\mathrm{PW}}(\rho, i\omega)
	= \frac{1}{2} \langle 2 k_x \mathrm{Tr}[{(i \hat{\sigma}_\nu \hat{\sigma}_2)^\dagger} 
	\hat{f}(\rho, \bs{k}, i \omega_n)
	] \rangle.
	\label{}
\end{align}

The free-energy density $\mathcal{F}(\bs{r})$ can be calculated from
the Green's function as,\cite{eilenberger:zphys1966,suzuki:prb2015,shu:prb2022}
\begin{align}
  & \mathcal{F} 
  = \mathcal{F}_f + \mathcal{F}_g, 
	\\
	& \mathcal{F}_f 
	= \pi N_0 T \sum_{\omega_n}
	\langle 
    \Delta^*(\bs{r})
    f(\bs{r},\bs{k},i \omega_n)
	\rangle, 
	\\
	& \mathcal{F}_g
	= 4 \pi N_0 T \sum_{\omega_n>0} ^{\omega_c}
	\int_{\omega_n}^{\omega_{c}'}
	\langle 
	  \mathrm{Re}[g(\bs{r},\bs{k},i \omega_n)-1 ]
  \rangle,
\label{}
\end{align}
where the free-energy density is measured from its normal
value; $\mathcal{F}=\mathcal{F}_S-\mathcal{F}_N$. 
In a homogeneous SC, the free-energy density approaches to
$\mathcal{F}(\rho) \to -\bar{\mathcal{F}}$ at low temperature with 
$\bar{\mathcal{F}}=N_0 \bar{\Delta}^2/2$ being the condensation energy
in the bulk. 

In the numerical simulations, we fix the parameters: 
$\omega_c = 6 \pi T_c$, $T=0.2T_c$, $\delta=0.01 \bar{\Delta}$, 
and $\omega'_c = 100 \bar{\Delta}$ with $\xi_0 = \hbar v_F / 2 \pi
T_c$ being the coherence length.

\section{Numerical results}
\subsection{ Pair potential and free-energy density}

We first discuss the influences of the exchange potential on the pair
potential as shown in Fig.~\ref{fig:del_Vdep}(a-c), where the
spatial variations of the 
pair potential are plotted for 
$V_0 /\bar{\Delta}=1$ in (a), $2$ in (b), and $3$ in (c).
The size of the magnetic cluster is $R_0 =3\xi_0$.  
The pair potential is almost homogeneous when the magnetization is
comparable to (or smaller than) the superconducting gap (i.e.,
$V_0<\bar{\Delta}$) as shown in Fig.~\ref{fig:del_Vdep}(a). 
We refer to {this} state as the zero-node state.  
For $V_0=2 \bar{\Delta}$ {[Fig.~\ref{fig:del_Vdep}(b)]}, the pair potential is
suppressed and changes its sign once around $\rho=2\xi_0$ (one-node state).  
For $V = 3 \bar{\Delta}$ {[Fig.~\ref{fig:del_Vdep}(c)]}, the pair potential changes the sign
twice (two-node state): negative approximately at
$1 < \rho/\xi_0 < 2.1$ and positive at $\rho/\xi_0<1$. 
Namely, the FFLO-like superconducting states are realized 
locally (i.e., only beneath the magnetic cluster).  

{
In the presence of the Zeeman field, 
the condition for the uniform superconducting state is given by $V_0< \bar{\Delta} /
\sqrt{2}$.\cite{chandrasekhar:apl1962,clogston:prl1962} 
}
The
superconducting state for $V_0=\bar{\Delta}$ in
Fig.~\ref{fig:del_Vdep}(a), however, goes {beyond} this limit.  
The pair potentials in Fig.~\ref{fig:del_Vdep}(a-c) are
{self-consistently}
obtained as stable solutions of the Eilenberger
equation. 
Such local FFLO states can be supported by the wide superconducting region
outside the cluster.
To confirm the validity of this argument, we calculate the free-energy density
$\mathcal{F}(\rho)$ 
in Fig.~\ref{fig:del_Vdep}(a-c).
The vertical axis is normalized to the condensation
energy in an uniform superconductor $\bar{\mathcal{F}}=\bar{\Delta}^2
N_0/2$ at zero temperature.  
The free-energy density outside the magnetic segment is negative 
(smaller than the free-energy density in the normal state)
and approaches to $-\bar{\mathcal{F}}$ for $\rho \gg R_0$.
Inside the magnetic segment, on the other hand, the free-energy density 
becomes positive locally.
 In particular, the free-energy density at $V_0 =
\bar{\Delta}$ in Fig.~\ref{fig:del_Vdep}(a) is always positive at
$\rho<R_0$.  However, the total free-energy
$\mathcal{F}_{\mathrm{Tot}} = \int \mathcal{F} d\bs{r}$ {is} always
negative because of the massive superconducting region outside the
cluster. 
Figure~\ref{fig:del_Vdep}(b) and \ref{fig:del_Vdep}(c) show that
introducing the nodes in the pair potential reduces the free-energy
density at $\rho <R_0$ drastically.  
Even so, the free energy at the magnetic segment 
$\mathcal{F}_{in}=\int_{r <R_0} d\boldsymbol{r} \mathcal{F}$ remains positive in 
the local FFLO states.
At $V_0 = 2\bar{\Delta}$ in Fig.~\ref{fig:del_Vdep}(b), 
$\mathcal{F}$ has a dip at $\rho \sim 1.9 \xi_0$ that corresponds to
the place of the node in the pair potential.
A similar behavior {appears} also in the results
for $V_0 = 3\bar{\Delta}$ {[Fig.~\ref{fig:del_Vdep}(c)]}. 
{The nodes in the pair potential and the dips in the free-energy
density appear almost at the same place. This correspondence, however,
contradicts intuition: }
The free-energy density {seems have} a peak (local maximum) around 
a node 
because the quasiparticle excitations below $\bar{\Delta}$ are allowed.  
The results in Figs~\ref{fig:del_Vdep}(b) and \ref{fig:del_Vdep}(c),
however, show the opposite tendency.
{
In Section~\ref{subsec:pair}, we will explore this further and discuss
how $p$-wave Cooper pairs localize at the nodes of the $s$-wave pair
potential, influencing the behavior of the free-energy density.
}

\subsection{Local density of states}
The {signatures} of the local FFLO state are accessible
through the LDOS of a quasiparticle which can be measured by the
scanning tunnel spectroscopy (STS) technique.  The numerical results
of the LDOS are shown in Fig.~\ref{fig:del_Vdep}(d-f), where the
exchange potentials are $V_0 /\bar{\Delta}=1$ in (d), 2 in (e), and 3
in (f).  The contour plots are shown as a function of the energy of a
quasiparticle $E$ and the radius $\rho$.  

At $V_0=\bar{\Delta}$, the
pair potential is almost homogeneous as shown in
Fig.~\ref{fig:del_Vdep}(a).  The LDOS in the ferromagnetic segment
indicates the appearance of quasiparticle states below the gap.  The
exchange potential pushes the coherence peak down to the subgap region
($|E|<\bar{\Delta}$) and broadens it in energy.  As a result, the LDOS
is slightly enhanced around $E=0.7 \bar{\Delta}$.  These states are
relating to the Yu-Shiba-Rusinov state\cite{yu:actphys1965,
shiba:ptp1968, rusinov:jetplett1969} localized around a point-like
magnetic impurity.\cite{shu:prb2022} The coherence peak at
$|E|=\bar{\Delta}$ can be found outside the magnetic segment
$\rho>R_0$.

When the exchange potential increases to $V_0=2\bar{\Delta}$ in (e),
the LDOS spectra show a complicated profile due to the spatial
variation of the pair potential.  The sharp subgap peak around $E= 0.7
\bar{\Delta}$ exist for $0 < \rho< 1.8\xi_0$ and $\rho>2.3 \xi_0$.  At
the node of the pair potential $\rho=2\xi_0$, the subgap peaks around
$E= 0.7\bar{\Delta}$ are drastically suppressed and the LDOS shows
almost flat spectra as it does in the normal state. 

The same tendency can be also seen in the results for $V_0=3
\bar{\Delta}$ in Fig.~\ref{fig:del_Vdep}(f). The spectra of LDOS for
$V_0=3 \bar{\Delta}$ in (f) become more inhomogeneous and complicated
than those in {Figs.~\ref{fig:del_Vdep}(d) and
Fig.~\ref{fig:del_Vdep}(e)}.  At the outer node at $\rho= 2.2 \xi_0$,
the spectra are totally flat and LDOS does not have large peaks below
the gap.  At the inner node at $\rho= \xi_0$, however, the LDOS has a
peak at zero energy.  
{In Sec.~\ref{Sec:1d}, we will discuss the
LDOS spectra are very sensitive to $V_0$ using the one-dimensional SF
structure. 
}

\subsection{Pairing correlations}\label{subsec:pair}
We display the pairing correlation functions in Figs.~\ref{fig:del_Vdep}(g-i)
for $V_0= \bar{\Delta}$, $2\bar{\Delta}$, and $3 \bar{\Delta}$, respectively.
 The results are calculated for the lowest frequency $\omega_0=\pi T$.   
 The spin-singlet $s$-wave component for $V_0= \bar{\Delta}$
is always larger than the other components and almost flat as shown in
Figs.~\ref{fig:del_Vdep}(g).  Only this component has a finite amplitude far
from the magnetic cluster. The two $p$-wave components show a broad
peak at the boundary ($\rho=R_0$) as a result of 
{the local inversion symmetry breaking}.
The odd-frequency triplet $s$-wave component has a relatively large 
amplitude than the induced $p$-wave components around the center.  

It is possible to derive the Eilenberger equation for corresponding 
four coherence functions: $\gamma_0^{\mathrm{SW}}$, $\gamma_0^{\mathrm{PW}}$, 
$\gamma_3^{\mathrm{SW}}$, $\gamma_3^{\mathrm{PW}}$.
The {detailed} results are displayed in the Appendix.
The equation {following} the first row of Eq.~(\ref{eq:el_four}),
\begin{align}
v_F\, k_x\, \frac{d}{dx} \gamma_0^{\mathrm{SW}} + 2\omega_n  \gamma_0^{\mathrm{PW}} + 2V 
\gamma_3^{\mathrm{PW}}=0, \label{eq:eilen_four1}
\end{align}
indicates that the spatial variation of the pair potential
{(spin-singlet $s$-wave superconductor)}
generates two $p$-wave components
$\gamma_0^{\mathrm{PW}}$ and $\gamma_3^{\mathrm{PW}}$.
In addition, 
the equation {following} the second row of Eq.~(\ref{eq:el_four}),
\begin{align}
v_F\, k_x\, \frac{d}{dx} \gamma_0^{\mathrm{PW}} + 2\omega_n  \gamma_0^{\mathrm{SW}} + 2V 
\gamma_3^{\mathrm{SW}}=\Delta,
\end{align}
explains the appearance of the spin-triplet $s$-wave component $\gamma_3^{\mathrm{SW}}$ 
even in a uniform pair potential.
Here we summarize our knowledge of the relation between the frequency symmetry of a Cooper pair
and their influence on the free energy and on the quasiparticle LDOS. 
\begin{enumerate}[I]
\item Usual \textit{even-frequency} pairs indicate the diamagnetic response to
magnetic fields and favor the spatially uniform superconducting phase
at the ground state.  On the other hand, \textit{odd-frequency} Cooper pairs
are paramagnetic.\cite{asano:prl2011} Therefore, odd-frequency pairs
increase the free energy of uniform ground state\cite{asano:prb2015}
and favor the spatial gradient of superconducting
phase.\cite{shu:prb2022} 

\item The LDOS has a gapped energy spectrum in the presence of even-frequency pairs, whereas it tends to have peaks below the gap in the presence of odd-frequency pairs. \cite{tanaka:prl2007,kim:jpsj2021}
\end{enumerate}

These properties qualitatively explain the characteristic behavior in
the free-energy density and those in the LDOS.
The LDOS for $V_0= \bar{\Delta}$ in Fig.~\ref{fig:del_Vdep}(d) shows the gap-like 
energy spectra because 
even-frequency component $f_0^{\mathrm{SW}}$ is dominant everywhere.
When {the exchange
potential increases} to
$V_0 = 2 \bar{\Delta}$, a node appears in the pair potential.  
As a consequence, the spatial profile of the pairing
correlations drastically changes as shown in Fig.~\ref{fig:del_Vdep}(h).
The two $p$-wave components have peaks inside the boundary because the
pair potential {and} the exchange potential break
the inversion symmetry locally. 
The amplitudes of the two $p$-wave components are larger than those 
in {Fig.~\ref{fig:del_Vdep}(g)}. 
The two spin-triplet components have large amplitudes around 
the node of the pair potential.
The sign change of the pair potential is equivalent to the local $\pi$-phase shift
in the pair potential.
According to the property I, odd-frequency spin-triplet $s$-wave components $f_3^{\mathrm{SW}}$ appear 
to decrease the free-energy density at the node of the pair potential.
The triplet $p$-wave component $f_3^{\mathrm{PW}}$ is the most dominant 
at the node $\rho=2\xi_0$. As a consequence, LDOS at the node in
{{Fig.~\ref{fig:del_Vdep}(e)}} does not have large peaks below the gap.
The two odd-frequency components are the source of the subgap peak at
$E=0.7\bar{\Delta}$ in {{Fig.~\ref{fig:del_Vdep}(e)}} for $0 <\rho< 1.8 \xi_0$.
Outside the magnetic segment, $f_0^{\mathrm{SW}}$ is a source of the
coherence peak at $E=\bar{\Delta}$ and $f_0^{\mathrm{PW}}$ assists the subgap peak at $E=0.7\bar{\Delta}$.
The two components $f_0^{\mathrm{SW}}$ and $f_3^{\mathrm{PW}}$ seem to affect the LDOS 
independently at $\rho >R_0$.

\begin{figure}[t]
	\includegraphics[width=0.44\textwidth]{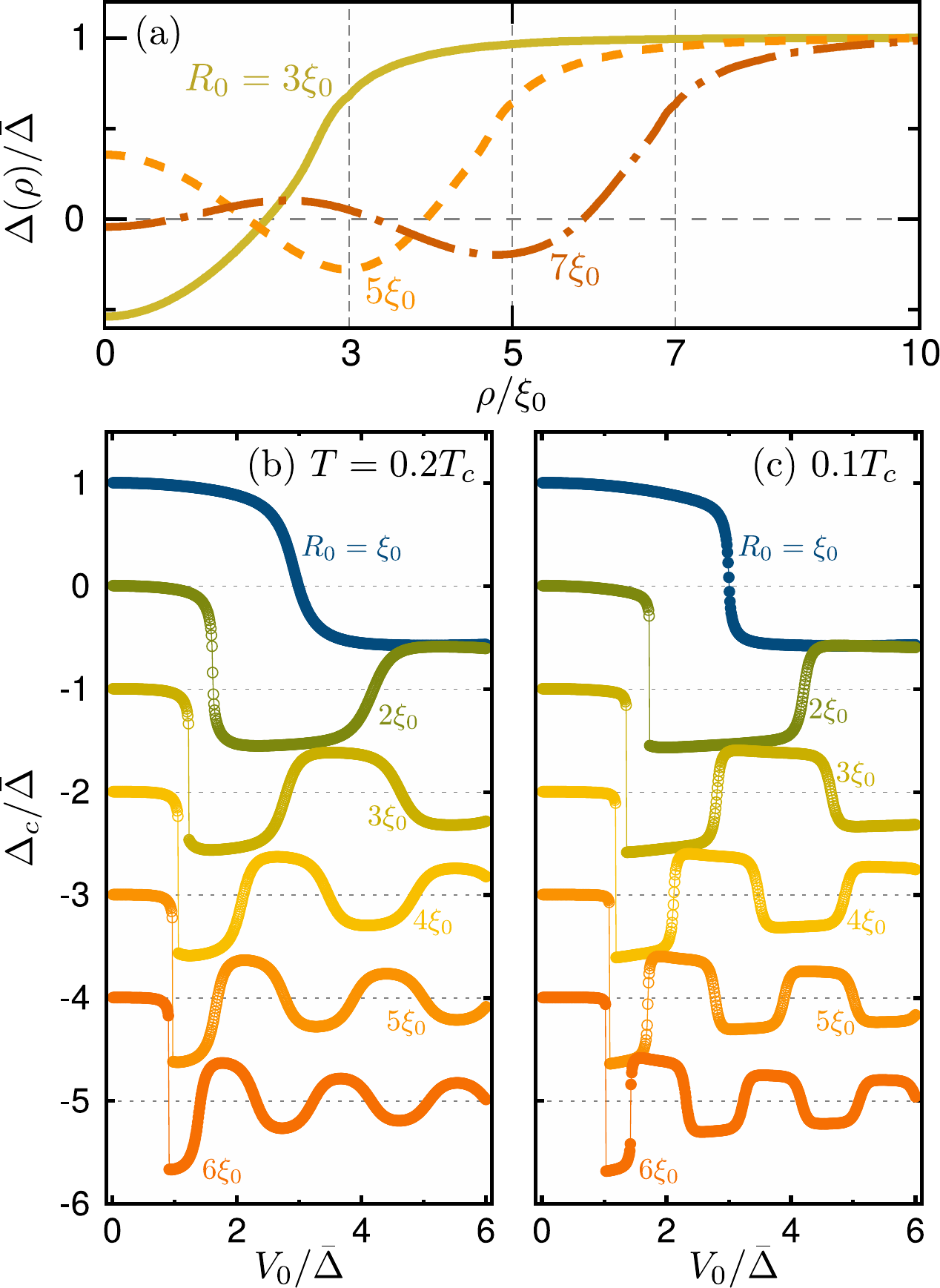}
	\caption{
	(a) Cluster-size dependence of pair potentials at $V_0= 2 \bar{\Delta}$.
	(b,c) 
	Pair potential at the center of the magnetic cluster ($\rho=0$)
	as a function of the exchange potential $V_0$. 
%
		The radius of the cluster varies from $R=\xi_0$ to $6\xi_0$ by
		$\xi_0$. The results are plotted with the offset by 
		$(R_0/\xi_0-1)\bar{\Delta}$ The horizontal broken lines indicate zeros. 
	  The temperature is set to (a,b) $T=0.2T_c$ 
		and (c) $0.1T_c$. 
  }
  \label{fig:del_delc_PD}
\end{figure}

In a two-node state at $V_0=3\bar{\Delta}$, 
the pairing correlation functions oscillate in the ferromagnetic segment more rapidly as shown in 
in Fig.~\ref{fig:del_Vdep}(i).
At the outer node $\rho = 2.2\xi_0$, the two spin-triplet components ($f_3^{\mathrm{PW}}$ and $f_3^{\mathrm{SW}}$) 
{have peaks}.   
The spectra of LDOS show the flat structure because the amplitude of
$f_3^{\mathrm{PW}}$ and that of $f_3^{\mathrm{SW}}$ are almost the
same at the outer node.  Around the inner node {at} $\rho = \xi_0$, the two
odd-frequency pairing correlations {have larger
amplitudes than} the two even-frequency
pairing correlations.  As a consequence, LDOS has a peak at zero
energy for $0.7\xi_0 < \rho < 2\xi_0$.  Therefore, the relative
amplitudes among the four correlation functions govern the subgap
spectra in the LDOS.  When the odd-frequency (even-frequency) pairing
correlations are dominant, the LDOS tend to have peak (gap) at
$E<\bar{\Delta}$. 

\begin{figure}[tb]
	\includegraphics[width=0.46\textwidth]{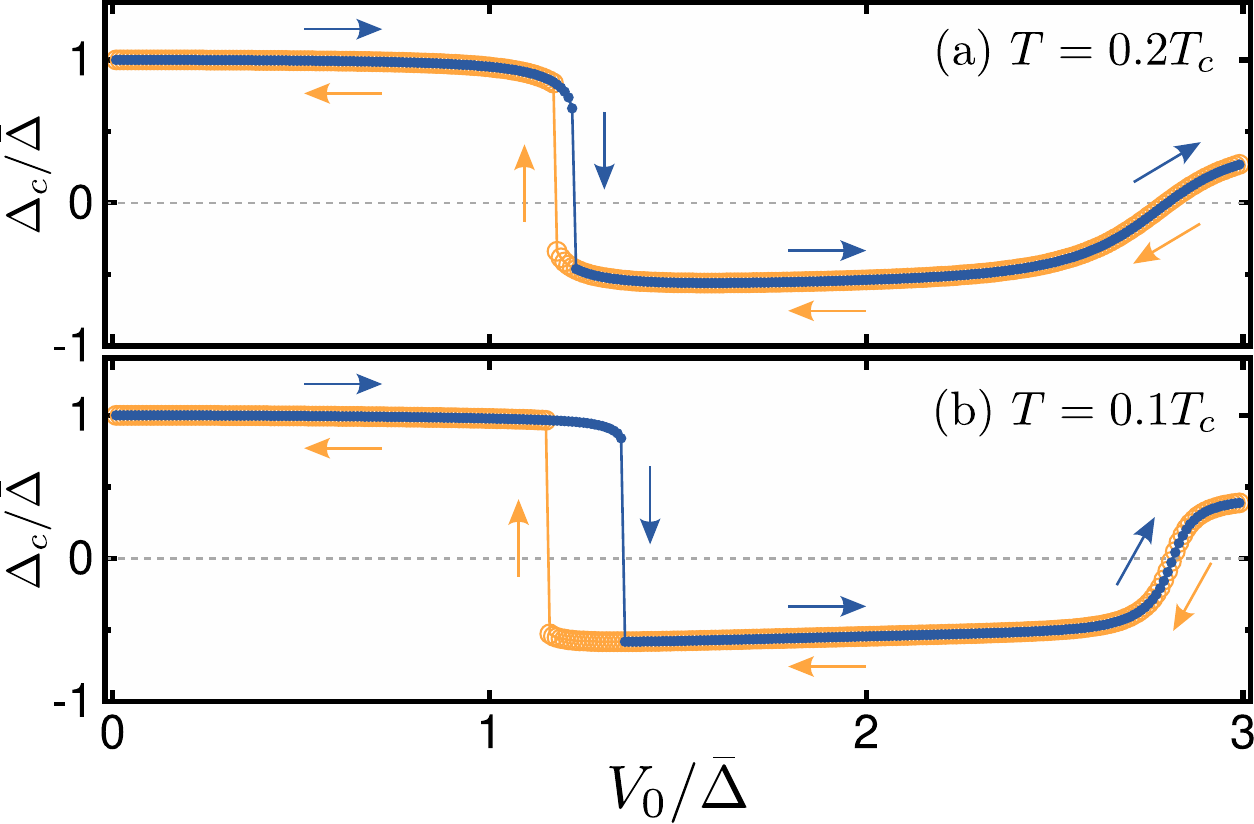}
	\caption{Hysteresis loop of $\Delta_c$. 
	The degree of the hysteresis is more prominent for larger magnetic clusters
	and at lower temperatures. 
	The arrows indicate how the exchange potential is changed in the numerical simulation.
	The radius of the island is set to $R_0=3\xi_0$.}
  \label{fig:del_delc_PD2}
\end{figure}

At the end of this subsection, we {briefly summarize}
the two properties of the local FFLO states. 
First, zero-energy peaks appear at the edge of the
ferromagnetic segment $\rho=R_0$ in Figs.~\ref{fig:del_Vdep}(e) and
\ref{fig:del_Vdep}(f). When $V_0$ is close to a transition
point between the $n$-node and $n\pm 1$-node states, the LDOS tends to have a
zero-energy peak at the boundary. 
However, this zero-energy peak is not universal and depends on the amplitude of the magnetization. We will further explore this issue in
Sec.~\ref{Sec:1d}.

Secondly, 
Eq.~(\ref{eq:eilen_four1}) implies that the spatial variation of the
singlet $s$-wave component (linked to the pair potential) generates
two $p$-wave components.  Simultaneously, we can state that
the $p$-wave pairing correlations drive the spatial variation of the
pair potential. Therefore, $p$-wave pairing correlations are
indispensable to realizing the FFLO states.
This insight consistent with a fact that the FFLO states are fragile
againt impurity scatterings.
\cite{aslamazov1969influence,
takada1970superconductivity,
PhysRevB.74.144522} 
To our knowledge, a spin-singlet $s$-wave
superconducting state is fragile when it contains odd-frequency
pairing correlations in the clean
limit.\cite{asano:prb2018,takumi:prb2020}

\subsection{Discontinuous transition}\label{subsec:disconti}

The spatial profiles of the pair potential with several radii of
the magnetic cluster are shown in Fig.~\ref{fig:del_delc_PD}(a), 
where we choose $R_0/\xi_0$= 3, 5, and 7 and $V_0 = 2 \bar{\Delta}$. 
The results indicate that, if the radius of the magnetic cluster is
large enough, the multi-node states appear even with a weak
exchange potential.  
In Figs.~\ref{fig:del_delc_PD}(b) and \ref{fig:del_delc_PD}(c), we plot the pair potential at the
center of the ferromagnetic segment $\Delta_c \equiv \Delta(\rho=0)$
as a function of $V_0$,\footnote{Changing $V_0$ in the horizontal axis
is realized by applying an external Zeeman field in addition to the
magnetic moment possessed in a ferromagnet.}
 where
the radius varies from $R_0 = \xi_0$ to $6\xi_0$ by $\xi_0$ with the
corresponding offset and the temperature is set to (b) $T/T_c =
0.2$ and (c) 0.1. 
The pair potential keeps the
homogeneous profile (i.e., $\Delta_c \approx \bar{\Delta}$ without a
node) until $V_0$ reaches to a critical value, which we define $V_1$.
At $V_0=V_1$, $\Delta_c$ changes the sign abruptly, meaning that a
node appears in the pair potential. 
Each time $\Delta_c$ changes the sign in Fig.~\ref{fig:del_delc_PD}, 
the number of nodes in the pair potential changes by one.
The results show a relation $V_1 \approx \bar{\Delta}$ holds
$R_0>2\xi_0$ which limits the validity of the theoretical model for 
a topologically nontrivial superconducting nanowire.\cite{oreg:prl2010,lutchyn:prl2010}

We also find jumps in $\Delta_c$ at $V_0=V_1$ shows the hysteresis
between the two processes in the numerical simulation: increasing and decreasing $V_0$.
The results for $R_0=3\xi_0$ are displayed in Fig.~\ref{fig:del_delc_PD2},
where we choose (a) $T=0.2T_c$ and (b) $0.1T_c$.  
The hysteresis loop is more prominent at a lower temperature. 
We have confirmed that the hysteresis loop
appears also between the one- and two-node states if the temperature is
low enough.

\begin{figure}[tb]
	\includegraphics[width=0.46\textwidth]{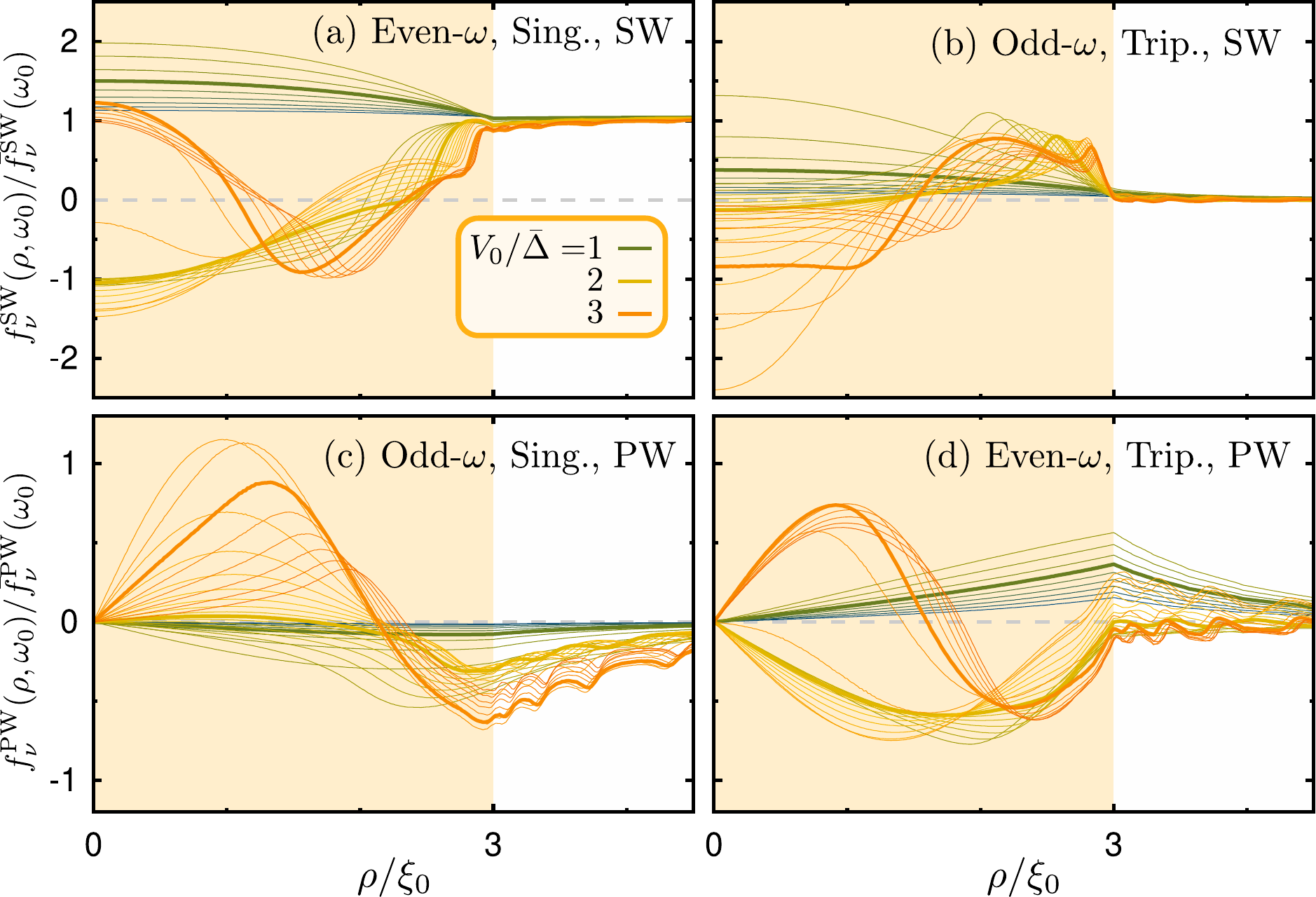}
	\caption{Evolutions of pair amplitudes over magnetization. 
			(a) Even-frequency spin-singlet $s$-wave, 
			(b)  Odd-frequency triplet $s$-wave, 
			(c)  Odd-frequency singlet $p$-wave, and 
			(d) Even-frequency triplet $p$-wave are shown. 
			The spatial profile of the pair potential is qualitatively the
			same as those in (a); the principal component. 
			The magnetization varies from
			$V_0/\bar{\Delta}=0.5$ to $4.0$ by
			$0.1$. The radius of the island and the temperature are set to
	$R_0=3\xi_0$ and $T=0.1T_c$.
		The even-frequency components (a,d) exhibit discrete behavior,
	while the odd-frequency components (b,c) vary gradually.
    The values of $V_0$ are given in the legend only for the thick lines.
	}
  \label{fig:pair}
\end{figure}

%
Comparing Figs.~\ref{fig:del_delc_PD}(b) and \ref{fig:del_delc_PD}(c), 
we see the discontinuous behavior is the more remarkable at the lower temperature.
The first term of the Eilenberger equation \eqref{eq:Riccati03} is 
$ [\partial_x + (2\xi_T)^{-1}] \hat{\gamma}$ 
along the $x$ coordinate, where we focus on the lowest Matsubara frequency 
and $\xi_T = \hbar v_F / 2 \pi T$ is the thermal coherence length in the clean limit. 
In this case, the length scale of the spatial variation of
$\hat{\gamma}$ is approximately given by $\xi_T$. 
In other words, the spatial variation in the coherence 
functions are correlated within the range with $\pi \xi_T^2$.
At $T=0.1T_c$, the thermal coherence length becomes $\xi_T=10 \xi_0$
which covers the whole ferromagnetic segment. 
In this case, $\Delta$ prefers a homogeneous profile as much as possible until $V_0$
exceeds a critical value because the influence of a node in the pair
potential spreads over a wide region of approximately $\pi \xi_T^2$. 
Therefore, at low temperature, the jump in 
$\Delta_c$ is more abrupt. 
Figure ~\ref{fig:del_delc_PD} also indicates that the discontinuity
between the zero-node and the one-node states is more remarkable for
larger clusters. At present, however, we cannot think of reasons for
this tendency.

In Fig.~\ref{fig:pair}, we display the pairing correlations functions
by changing the exchange potential gradually for $R_0=3\xi_0$.  The
spatial profile of the spin-singlet $s$-wave component in the $n$-node
state is qualitatively different from those in $n\pm 1$-node states
[Fig.~\ref{fig:pair}(a)]. Roughly speaking, the spatial
profiles of $f_0^{\mathrm{SW}}$ is insensitive to $V_0$ as long as the
number of nodes in the pair potential remains the same value.  As a
result, the calculated results for $f_0^{\mathrm{SW}}$ are bundled for
each {state}. The same discontinuous behavior
appears in the even-frequency spin-triplet $p$-wave component in
Fig.~\ref{fig:pair}(d).  Such discontinuous behavior in the pairing
correlation functions is responsible for the jump of the pair
potential between the $n$-node state and $n +1$-node state.  The
spatial profile of the two even-frequency pairing components is
governed mainly by the number of nodes.  The two odd-frequency
components in Fig.~\ref{fig:pair}(b) and \ref{fig:pair}(c), on the other hand,
changes gradually with increasing $V_0$.  Consequently, odd-frequency
pairs relax the effects of discontinuous change in the pair potential
at the transition point.  This might be a role of odd-frequency pairs
in the FFLO states.  The gradual change of the odd-frequency pairing
correlations causes the gradual change of subgap spectra in the LDOS.
We will discuss this issue in Sec.~\ref{Sec:1d}.

\section{Analysis in one-channel model}\label{Sec:1d}

\begin{figure}[t]
\centering
\includegraphics[width=0.48\textwidth]{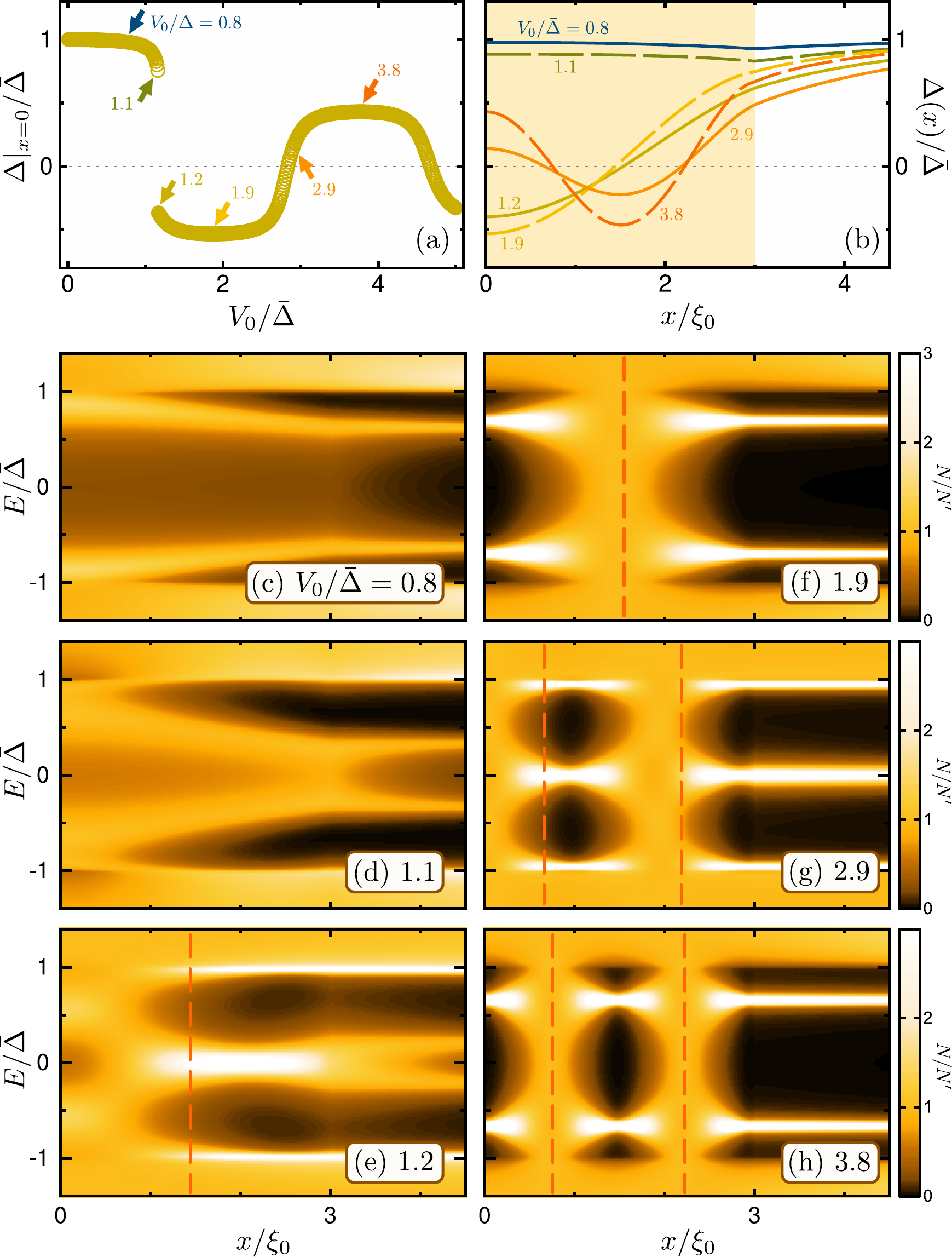}
\caption{Results for the one-channel model. 
The length of the ferromagnet and the temperature are set to $L_0=3 \xi_0$ and
$T=0.2T_c$, respectively. 
(a) Pair potential at the center of the system. 
(b) Spatial profiles of the pair potential. 
(c-h) Local density of states. 
The magnetization is set to the
characteristic values indicated by the arrows in (a): 
$V_0/\bar{\Delta} = 0.8$ (zero-node), 
$1.1$ (before the transition), 
$1.2$ (after the transition), 
$1.9$ (one-node state), 
$2.9$ (after the second transition), and 
$3.8$ (two-node state). 
The vertical broken lines in (c-h) indicate the position of the nodes. }
\label{fig:dos_1d}
\end{figure}

The one-channel model is represented by putting $k_y=0$ in all the 
equations in Sec.~\ref{sec:model} and describes a one-dimensional superconducting 
structure including the exchange potential of $V(x) = V_0
{\Theta(L_0-|x|)}$. 
Although the one-channel model is not realistic, 
the characteristic behaviors in the LDOS in the one-channel model are
simpler than those in the two-dimension.

The pair potential at the center of the system (i.e.,
$\Delta_c=\Delta|_{x=0}$ is shown in
Fig.~\ref{fig:dos_1d}(a) where $L_0=3\xi_0$ and $T=0.2T_c$. We focus
on the several characteristic exchange potentials indicated by the arrows: 
$V_0/\bar{\Delta} = 0.8$ (zero-node), 
$1.1$ (just before the transition), 
$1.2$ (just after the transition), 
$1.9$ (one-node state), 
$2.9$ ({just} after the second transition), and 
$3.8$ (two-node state). The spatial profile of the pair potentials
in Fig.~\ref{fig:dos_1d}(b) is qualitatively the same as those in 
the 2D case [See Fig.~\ref{fig:del_Vdep}(a-c)]. 
We have also confirmed that the transition between the one-node and two-node states
becomes discontinuous at a low temperature {(The results are
not shown)}. 
The LDOS for {the characteristic} exchange potentials 
are shown in
Fig.~\ref{fig:dos_1d}(c-h). The LDOS indicate that the quasiparticle spectra are
not simply determined by the number of nodes but depend sensitively on 
the exchange potential. 
In the zero-node state in Fig.~\ref{fig:dos_1d}(c),  
the LDOS at the interface ($x=3\xi_0$) has peaks around $|E| = 0.6$ and
that at the center of the magnetic segment has peaks around $|E| = 0.9$.  
Just below the transition point to the one-node state, the LDOS  
at the boundary ($x=3\xi_0$) has a peak at zero energy as shown in Fig.~\ref{fig:dos_1d}(d).
The zero-energy peak at the boundary can be seen also just above the transition point 
in Fig.~\ref{fig:dos_1d}(e) and \ref{fig:dos_1d}(g).
At $V_0/\bar{\Delta} = 1.9$, the one-node state is stable 
because the exchange potential is far from the two transition points of 
$V_0/\bar{\Delta} = 1.15$ and $2.85$ [see Fig.~\ref{fig:dos_1d}(a)].
The corresponding LDOS [Fig.~\ref{fig:dos_1d}(f)] shows that the spectra are flat 
$N \approx N_0$ at the node of the pair potential and gapped at the boundary.
The same behavior appears also for the stable two-node states at $V_0/\bar{\Delta} = 3.8$
[Fig.~\ref{fig:dos_1d}(h)]. 
	Although the pair potentials for $V_0/\bar{\Delta} = 1.2$ and $1.9$
	have the similar profiles, 
	the subgap spectra [Figs.~\ref{fig:dos_1d}(e) and
	\ref{fig:dos_1d}(f)] are qualitatively different. 
As already discussed in Fig.~\ref{fig:pair}, the spatial profiles of
the odd-frequency components for these states deviate from each other.
As a result, the subgap spectra in (e) and (f) are totally different
from each other according to property II. 
The spatial profiles of
the odd-frequency components are very sensitive to $V_0$ and those 
for these two states are qualitatively
different (the results are not shown but similar to those in Fig.~\ref{fig:pair}).
Therefore, the subgap spectra shown in Figs.~\ref{fig:dos_1d}(e) and \ref{fig:dos_1d}(f) exhibit distinct differences, which can be attributed to property II.
This discussion can be
applied also to the LDOS in the two two-node states shown in
Figs.~\ref{fig:dos_1d}(g) and \ref{fig:dos_1d}(h).  Thus, the gradual changes of the
odd-frequency pairing correlations are responsible for the gradual
changes in the LDOS.

\section{Conclusion}

We have theoretically studied the property of the
{Fulde-Ferrell-Larkin-Ovchinnikov (FFLO)} state 
which appears in a superconducting thin film attached to a circular-shaped magnetic cluster.  
By solving the quasiclassical Eilenberger equation, we calculate the pair potential,
free-energy density, the pairing correlation functions, and the local density of states.
The FFLO states are locally realized beneath the ferromagnetic cluster as a
stable solution of the self-consistent gap equation.  
The free-energy density shows that the
local FFLO states are supported by superconducting condensate
surrounding the magnetic cluster.  
{As the exchange
potential increases}, the number of nodes in the pair potential 
increases one by one.

The spatial profiles of the even-frequency pairing correlations are 
not sensitive to the exchange potential but are determined 
mainly by the number of nodes in the pair potential.
On the other hand, the odd-frequency pairing 
correlations show a  gradual change with increasing the exchange potential.
The local density of states is inhomogeneous in the {local} FFLO state.
In addition, the subgap spectra depend sensitively on the exchange potential
 because the odd-frequency pairs coexist with subgap quasiparticles.

\begin{acknowledgments}
This work was supported by JSPS KAKENHI (No. JP20H01857). 
S.-I.~S. acknowledges Overseas Research Fellowships by JSPS and the hospitality at the University of Twente.
T.~S. is supported in part by the establishment of university 
fellowships towards the creation of science technology innovation from the 
Ministry of Education, Culture, Sports, Science, and Technology (MEXT) of Japan.
\end{acknowledgments}


\appendix*
\section{Analysis in linearized Eilenberger equation}\label{Sec:le}
It is possible to derive the
 Eilenberger equation for the corresponding four coherence functions defined by
 \begin{align}
\gamma_0^{\mathrm{SW}}=& \frac{1}{2}\left\{\gamma_0(\boldsymbol{k}) + \gamma_0(-\boldsymbol{k}) \right\},\\
\gamma_0^{\mathrm{PW}}=& \frac{1}{2}\left\{\gamma_0(\boldsymbol{k}) - \gamma_0(-\boldsymbol{k}) \right\},\\
\gamma_3^{\mathrm{SW}}=& \frac{1}{2i}\left\{\gamma_3(\boldsymbol{k}) + \gamma_3(-\boldsymbol{k}) \right\},\\
\gamma_3^{\mathrm{PW}}=& \frac{1}{2i}\left\{\gamma_3(\boldsymbol{k}) - \gamma_3(-\boldsymbol{k}) \right\}.
\end{align}
Here we assume that an $s$-wave ($p$-wave) component is the most dominant in an even-parity (odd-parity) 
coherence function.
The Eilenberger equation for these components results in
\begin{align}
v_F \, \boldsymbol{k} \cdot \nabla&
\left[ \begin{array}{c} \gamma_0^{\mathrm{SW}} \\
\gamma_0^{\mathrm{PW}} \\
\gamma_3^{\mathrm{SW}} \\
\gamma_3^{\mathrm{PW}} 
\end{array}\right]
+2\left[ \begin{array}{cccc}
0 & \omega & 0& V\\
\omega & 0 & V & 0\\
0 & -V & 0 & \omega \\
-V & 0 & \omega& 0
\end{array}
\right]
\left[ \begin{array}{c} \gamma_0^{\mathrm{SW}} \\
\gamma_0^{\mathrm{PW}} \\
\gamma_3^{\mathrm{SW}} \\
\gamma_3^{\mathrm{PW}} 
\end{array}\right]
\nonumber\\
+&\left[ \begin{array}{c} 
2 ( \gamma_0^{\mathrm{SW}} \gamma_0^{\mathrm{PW}} - \gamma_3^{\mathrm{SW}} \gamma_3^{\mathrm{PW}}) \\
(\gamma_0^{\mathrm{SW}})^2 +(\gamma_0^{\mathrm{PW}})^2 - (\gamma_3^{\mathrm{SW}})^2 - (\gamma_3^{\mathrm{PW}})^2 \\
2 ( \gamma_0^{\mathrm{SW}} \gamma_3^{\mathrm{PW}} + \gamma_0^{\mathrm{PW}} \gamma_3^{\mathrm{SW}}) \\
2 ( \gamma_0^{\mathrm{SW}} \gamma_3^{\mathrm{SW}} - \gamma_0^{\mathrm{PW}} \gamma_3^{\mathrm{PW}}) 
\end{array}
\right] \nonumber\\
=&
\left[ \begin{array}{c} 0 \\
\Delta \\
0\\
0
\end{array}\right].\label{eq:el_four}
\end{align}
The pair potential is calculated from an spin-singlet $s$-wave component $\gamma_0^{\mathrm{SW}}$.

\bibliography{list_2023}

\begin{thebibliography}{70}%
\makeatletter
\providecommand \@ifxundefined [1]{%
 \@ifx{#1\undefined}
}%
\providecommand \@ifnum [1]{%
 \ifnum #1\expandafter \@firstoftwo
 \else \expandafter \@secondoftwo
 \fi
}%
\providecommand \@ifx [1]{%
 \ifx #1\expandafter \@firstoftwo
 \else \expandafter \@secondoftwo
 \fi
}%
\providecommand \natexlab [1]{#1}%
\providecommand \enquote  [1]{``#1''}%
\providecommand \bibnamefont  [1]{#1}%
\providecommand \bibfnamefont [1]{#1}%
\providecommand \citenamefont [1]{#1}%
\providecommand \href@noop [0]{\@secondoftwo}%
\providecommand \href [0]{\begingroup \@sanitize@url \@href}%
\providecommand \@href[1]{\@@startlink{#1}\@@href}%
\providecommand \@@href[1]{\endgroup#1\@@endlink}%
\providecommand \@sanitize@url [0]{\catcode `\\12\catcode `\$12\catcode
  `\&12\catcode `\#12\catcode `\^12\catcode `\_12\catcode `\%12\relax}%
\providecommand \@@startlink[1]{}%
\providecommand \@@endlink[0]{}%
\providecommand \url  [0]{\begingroup\@sanitize@url \@url }%
\providecommand \@url [1]{\endgroup\@href {#1}{\urlprefix }}%
\providecommand \urlprefix  [0]{URL }%
\providecommand \Eprint [0]{\href }%
\providecommand \doibase [0]{http://dx.doi.org/}%
\providecommand \selectlanguage [0]{\@gobble}%
\providecommand \bibinfo  [0]{\@secondoftwo}%
\providecommand \bibfield  [0]{\@secondoftwo}%
\providecommand \translation [1]{[#1]}%
\providecommand \BibitemOpen [0]{}%
\providecommand \bibitemStop [0]{}%
\providecommand \bibitemNoStop [0]{.\EOS\space}%
\providecommand \EOS [0]{\spacefactor3000\relax}%
\providecommand \BibitemShut  [1]{\csname bibitem#1\endcsname}%
\let\auto@bib@innerbib\@empty
\bibitem [{\citenamefont {Fulde}\ and\ \citenamefont
  {Ferrell}(1964)}]{fulde_fflo}%
  \BibitemOpen
  \bibfield  {author} {\bibinfo {author} {\bibfnamefont {P.}~\bibnamefont
  {Fulde}}\ and\ \bibinfo {author} {\bibfnamefont {R.~A.}\ \bibnamefont
  {Ferrell}},\ }\href {\doibase 10.1103/PhysRev.135.A550} {\bibfield  {journal}
  {\bibinfo  {journal} {Phys. Rev.}\ }\textbf {\bibinfo {volume} {135}},\
  \bibinfo {pages} {A550} (\bibinfo {year} {1964})}\BibitemShut {NoStop}%
\bibitem [{\citenamefont {Larkin}\ and\ \citenamefont
  {Ovchinnikov}(1965)}]{larkin_fflo}%
  \BibitemOpen
  \bibfield  {author} {\bibinfo {author} {\bibfnamefont {A.~I.}\ \bibnamefont
  {Larkin}}\ and\ \bibinfo {author} {\bibfnamefont {Y.~N.}\ \bibnamefont
  {Ovchinnikov}},\ }\href@noop {} {\bibfield  {journal} {\bibinfo  {journal}
  {Sov. Phys. JETP}\ }\textbf {\bibinfo {volume} {20}},\ \bibinfo {pages} {762}
  (\bibinfo {year} {1965})}\BibitemShut {NoStop}%
\bibitem [{\citenamefont {Lortz}\ \emph {et~al.}(2007)\citenamefont {Lortz},
  \citenamefont {Wang}, \citenamefont {Demuer}, \citenamefont {B\"ottger},
  \citenamefont {Bergk}, \citenamefont {Zwicknagl}, \citenamefont {Nakazawa},\
  and\ \citenamefont {Wosnitza}}]{PhysRevLett.99.187002}%
  \BibitemOpen
  \bibfield  {author} {\bibinfo {author} {\bibfnamefont {R.}~\bibnamefont
  {Lortz}}, \bibinfo {author} {\bibfnamefont {Y.}~\bibnamefont {Wang}},
  \bibinfo {author} {\bibfnamefont {A.}~\bibnamefont {Demuer}}, \bibinfo
  {author} {\bibfnamefont {P.~H.~M.}\ \bibnamefont {B\"ottger}}, \bibinfo
  {author} {\bibfnamefont {B.}~\bibnamefont {Bergk}}, \bibinfo {author}
  {\bibfnamefont {G.}~\bibnamefont {Zwicknagl}}, \bibinfo {author}
  {\bibfnamefont {Y.}~\bibnamefont {Nakazawa}}, \ and\ \bibinfo {author}
  {\bibfnamefont {J.}~\bibnamefont {Wosnitza}},\ }\href {\doibase
  10.1103/PhysRevLett.99.187002} {\bibfield  {journal} {\bibinfo  {journal}
  {Phys. Rev. Lett.}\ }\textbf {\bibinfo {volume} {99}},\ \bibinfo {pages}
  {187002} (\bibinfo {year} {2007})}\BibitemShut {NoStop}%
\bibitem [{\citenamefont {Yonezawa}\ \emph {et~al.}(2008)\citenamefont
  {Yonezawa}, \citenamefont {Kusaba}, \citenamefont {Maeno}, \citenamefont
  {Auban-Senzier}, \citenamefont {Pasquier}, \citenamefont {Bechgaard},\ and\
  \citenamefont {J\'erome}}]{PhysRevLett.100.117002}%
  \BibitemOpen
  \bibfield  {author} {\bibinfo {author} {\bibfnamefont {S.}~\bibnamefont
  {Yonezawa}}, \bibinfo {author} {\bibfnamefont {S.}~\bibnamefont {Kusaba}},
  \bibinfo {author} {\bibfnamefont {Y.}~\bibnamefont {Maeno}}, \bibinfo
  {author} {\bibfnamefont {P.}~\bibnamefont {Auban-Senzier}}, \bibinfo {author}
  {\bibfnamefont {C.}~\bibnamefont {Pasquier}}, \bibinfo {author}
  {\bibfnamefont {K.}~\bibnamefont {Bechgaard}}, \ and\ \bibinfo {author}
  {\bibfnamefont {D.}~\bibnamefont {J\'erome}},\ }\href {\doibase
  10.1103/PhysRevLett.100.117002} {\bibfield  {journal} {\bibinfo  {journal}
  {Phys. Rev. Lett.}\ }\textbf {\bibinfo {volume} {100}},\ \bibinfo {pages}
  {117002} (\bibinfo {year} {2008})}\BibitemShut {NoStop}%
\bibitem [{\citenamefont {Agosta}\ \emph {et~al.}(2012)\citenamefont {Agosta},
  \citenamefont {Jin}, \citenamefont {Coniglio}, \citenamefont {Smith},
  \citenamefont {Cho}, \citenamefont {Stroe}, \citenamefont {Martin},
  \citenamefont {Tozer}, \citenamefont {Murphy}, \citenamefont {Palm},
  \citenamefont {Schlueter},\ and\ \citenamefont
  {Kurmoo}}]{PhysRevB.85.214514}%
  \BibitemOpen
  \bibfield  {author} {\bibinfo {author} {\bibfnamefont {C.~C.}\ \bibnamefont
  {Agosta}}, \bibinfo {author} {\bibfnamefont {J.}~\bibnamefont {Jin}},
  \bibinfo {author} {\bibfnamefont {W.~A.}\ \bibnamefont {Coniglio}}, \bibinfo
  {author} {\bibfnamefont {B.~E.}\ \bibnamefont {Smith}}, \bibinfo {author}
  {\bibfnamefont {K.}~\bibnamefont {Cho}}, \bibinfo {author} {\bibfnamefont
  {I.}~\bibnamefont {Stroe}}, \bibinfo {author} {\bibfnamefont
  {C.}~\bibnamefont {Martin}}, \bibinfo {author} {\bibfnamefont {S.~W.}\
  \bibnamefont {Tozer}}, \bibinfo {author} {\bibfnamefont {T.~P.}\ \bibnamefont
  {Murphy}}, \bibinfo {author} {\bibfnamefont {E.~C.}\ \bibnamefont {Palm}},
  \bibinfo {author} {\bibfnamefont {J.~A.}\ \bibnamefont {Schlueter}}, \ and\
  \bibinfo {author} {\bibfnamefont {M.}~\bibnamefont {Kurmoo}},\ }\href
  {\doibase 10.1103/PhysRevB.85.214514} {\bibfield  {journal} {\bibinfo
  {journal} {Phys. Rev. B}\ }\textbf {\bibinfo {volume} {85}},\ \bibinfo
  {pages} {214514} (\bibinfo {year} {2012})}\BibitemShut {NoStop}%
\bibitem [{\citenamefont {Croitoru}\ and\ \citenamefont
  {Buzdin}(2012)}]{PhysRevB.86.064507}%
  \BibitemOpen
  \bibfield  {author} {\bibinfo {author} {\bibfnamefont {M.~D.}\ \bibnamefont
  {Croitoru}}\ and\ \bibinfo {author} {\bibfnamefont {A.~I.}\ \bibnamefont
  {Buzdin}},\ }\href {\doibase 10.1103/PhysRevB.86.064507} {\bibfield
  {journal} {\bibinfo  {journal} {Phys. Rev. B}\ }\textbf {\bibinfo {volume}
  {86}},\ \bibinfo {pages} {064507} (\bibinfo {year} {2012})}\BibitemShut
  {NoStop}%
\bibitem [{\citenamefont {Croitoru}\ and\ \citenamefont
  {Buzdin}(2013)}]{croitoru2013fulde}%
  \BibitemOpen
  \bibfield  {author} {\bibinfo {author} {\bibfnamefont {M.~D.}\ \bibnamefont
  {Croitoru}}\ and\ \bibinfo {author} {\bibfnamefont {A.~I.}\ \bibnamefont
  {Buzdin}},\ }\href@noop {} {\bibfield  {journal} {\bibinfo  {journal}
  {Journal of Physics: Condensed Matter}\ }\textbf {\bibinfo {volume} {25}},\
  \bibinfo {pages} {125702} (\bibinfo {year} {2013})}\BibitemShut {NoStop}%
\bibitem [{\citenamefont {Mayaffre}\ \emph {et~al.}(2014)\citenamefont
  {Mayaffre}, \citenamefont {Kr{\"a}mer}, \citenamefont {Horvati{\'c}},
  \citenamefont {Berthier}, \citenamefont {Miyagawa}, \citenamefont {Kanoda},\
  and\ \citenamefont {Mitrovi{\'c}}}]{mayaffre2014evidence}%
  \BibitemOpen
  \bibfield  {author} {\bibinfo {author} {\bibfnamefont {H.}~\bibnamefont
  {Mayaffre}}, \bibinfo {author} {\bibfnamefont {S.}~\bibnamefont
  {Kr{\"a}mer}}, \bibinfo {author} {\bibfnamefont {M.}~\bibnamefont
  {Horvati{\'c}}}, \bibinfo {author} {\bibfnamefont {C.}~\bibnamefont
  {Berthier}}, \bibinfo {author} {\bibfnamefont {K.}~\bibnamefont {Miyagawa}},
  \bibinfo {author} {\bibfnamefont {K.}~\bibnamefont {Kanoda}}, \ and\ \bibinfo
  {author} {\bibfnamefont {V.}~\bibnamefont {Mitrovi{\'c}}},\ }\href@noop {}
  {\bibfield  {journal} {\bibinfo  {journal} {Nature Physics}\ }\textbf
  {\bibinfo {volume} {10}},\ \bibinfo {pages} {928} (\bibinfo {year}
  {2014})}\BibitemShut {NoStop}%
\bibitem [{\citenamefont {Koutroulakis}\ \emph {et~al.}(2016)\citenamefont
  {Koutroulakis}, \citenamefont {K\"uhne}, \citenamefont {Schlueter},
  \citenamefont {Wosnitza},\ and\ \citenamefont
  {Brown}}]{PhysRevLett.116.067003}%
  \BibitemOpen
  \bibfield  {author} {\bibinfo {author} {\bibfnamefont {G.}~\bibnamefont
  {Koutroulakis}}, \bibinfo {author} {\bibfnamefont {H.}~\bibnamefont
  {K\"uhne}}, \bibinfo {author} {\bibfnamefont {J.~A.}\ \bibnamefont
  {Schlueter}}, \bibinfo {author} {\bibfnamefont {J.}~\bibnamefont {Wosnitza}},
  \ and\ \bibinfo {author} {\bibfnamefont {S.~E.}\ \bibnamefont {Brown}},\
  }\href {\doibase 10.1103/PhysRevLett.116.067003} {\bibfield  {journal}
  {\bibinfo  {journal} {Phys. Rev. Lett.}\ }\textbf {\bibinfo {volume} {116}},\
  \bibinfo {pages} {067003} (\bibinfo {year} {2016})}\BibitemShut {NoStop}%
\bibitem [{\citenamefont {Imajo}\ \emph {et~al.}(2021)\citenamefont {Imajo},
  \citenamefont {Kobayashi}, \citenamefont {Kawamoto}, \citenamefont {Kindo},\
  and\ \citenamefont {Nakazawa}}]{PhysRevB.103.L220501}%
  \BibitemOpen
  \bibfield  {author} {\bibinfo {author} {\bibfnamefont {S.}~\bibnamefont
  {Imajo}}, \bibinfo {author} {\bibfnamefont {T.}~\bibnamefont {Kobayashi}},
  \bibinfo {author} {\bibfnamefont {A.}~\bibnamefont {Kawamoto}}, \bibinfo
  {author} {\bibfnamefont {K.}~\bibnamefont {Kindo}}, \ and\ \bibinfo {author}
  {\bibfnamefont {Y.}~\bibnamefont {Nakazawa}},\ }\href {\doibase
  10.1103/PhysRevB.103.L220501} {\bibfield  {journal} {\bibinfo  {journal}
  {Phys. Rev. B}\ }\textbf {\bibinfo {volume} {103}},\ \bibinfo {pages}
  {L220501} (\bibinfo {year} {2021})}\BibitemShut {NoStop}%
\bibitem [{\citenamefont {Kasahara}\ \emph {et~al.}(2020)\citenamefont
  {Kasahara}, \citenamefont {Sato}, \citenamefont {Licciardello}, \citenamefont
  {\ifmmode~\check{C}\else \v{C}\fi{}ulo}, \citenamefont
  {Arsenijevi\ifmmode~\acute{c}\else \'{c}\fi{}}, \citenamefont {Ottenbros},
  \citenamefont {Tominaga}, \citenamefont {B\"oker}, \citenamefont {Eremin},
  \citenamefont {Shibauchi}, \citenamefont {Wosnitza}, \citenamefont {Hussey},\
  and\ \citenamefont {Matsuda}}]{kasahara:prl2020}%
  \BibitemOpen
  \bibfield  {author} {\bibinfo {author} {\bibfnamefont {S.}~\bibnamefont
  {Kasahara}}, \bibinfo {author} {\bibfnamefont {Y.}~\bibnamefont {Sato}},
  \bibinfo {author} {\bibfnamefont {S.}~\bibnamefont {Licciardello}}, \bibinfo
  {author} {\bibfnamefont {M.}~\bibnamefont {\ifmmode~\check{C}\else
  \v{C}\fi{}ulo}}, \bibinfo {author} {\bibfnamefont {S.}~\bibnamefont
  {Arsenijevi\ifmmode~\acute{c}\else \'{c}\fi{}}}, \bibinfo {author}
  {\bibfnamefont {T.}~\bibnamefont {Ottenbros}}, \bibinfo {author}
  {\bibfnamefont {T.}~\bibnamefont {Tominaga}}, \bibinfo {author}
  {\bibfnamefont {J.}~\bibnamefont {B\"oker}}, \bibinfo {author} {\bibfnamefont
  {I.}~\bibnamefont {Eremin}}, \bibinfo {author} {\bibfnamefont
  {T.}~\bibnamefont {Shibauchi}}, \bibinfo {author} {\bibfnamefont
  {J.}~\bibnamefont {Wosnitza}}, \bibinfo {author} {\bibfnamefont {N.~E.}\
  \bibnamefont {Hussey}}, \ and\ \bibinfo {author} {\bibfnamefont
  {Y.}~\bibnamefont {Matsuda}},\ }\href {\doibase
  10.1103/PhysRevLett.124.107001} {\bibfield  {journal} {\bibinfo  {journal}
  {Phys. Rev. Lett.}\ }\textbf {\bibinfo {volume} {124}},\ \bibinfo {pages}
  {107001} (\bibinfo {year} {2020})}\BibitemShut {NoStop}%
\bibitem [{\citenamefont {Kasahara}\ \emph {et~al.}(2021)\citenamefont
  {Kasahara}, \citenamefont {Suzuki}, \citenamefont {Machida}, \citenamefont
  {Sato}, \citenamefont {Ukai}, \citenamefont {Murayama}, \citenamefont
  {Suetsugu}, \citenamefont {Kasahara}, \citenamefont {Shibauchi},
  \citenamefont {Hanaguri},\ and\ \citenamefont {Matsuda}}]{kasahara:prl2021}%
  \BibitemOpen
  \bibfield  {author} {\bibinfo {author} {\bibfnamefont {S.}~\bibnamefont
  {Kasahara}}, \bibinfo {author} {\bibfnamefont {H.}~\bibnamefont {Suzuki}},
  \bibinfo {author} {\bibfnamefont {T.}~\bibnamefont {Machida}}, \bibinfo
  {author} {\bibfnamefont {Y.}~\bibnamefont {Sato}}, \bibinfo {author}
  {\bibfnamefont {Y.}~\bibnamefont {Ukai}}, \bibinfo {author} {\bibfnamefont
  {H.}~\bibnamefont {Murayama}}, \bibinfo {author} {\bibfnamefont
  {S.}~\bibnamefont {Suetsugu}}, \bibinfo {author} {\bibfnamefont
  {Y.}~\bibnamefont {Kasahara}}, \bibinfo {author} {\bibfnamefont
  {T.}~\bibnamefont {Shibauchi}}, \bibinfo {author} {\bibfnamefont
  {T.}~\bibnamefont {Hanaguri}}, \ and\ \bibinfo {author} {\bibfnamefont
  {Y.}~\bibnamefont {Matsuda}},\ }\href {\doibase
  10.1103/PhysRevLett.127.257001} {\bibfield  {journal} {\bibinfo  {journal}
  {Phys. Rev. Lett.}\ }\textbf {\bibinfo {volume} {127}},\ \bibinfo {pages}
  {257001} (\bibinfo {year} {2021})}\BibitemShut {NoStop}%
\bibitem [{\citenamefont {Kinjo}\ \emph {et~al.}(2022)\citenamefont {Kinjo},
  \citenamefont {Manago}, \citenamefont {Kitagawa}, \citenamefont {Mao},
  \citenamefont {Yonezawa}, \citenamefont {Maeno},\ and\ \citenamefont
  {Ishida}}]{kinjo2022superconducting}%
  \BibitemOpen
  \bibfield  {author} {\bibinfo {author} {\bibfnamefont {K.}~\bibnamefont
  {Kinjo}}, \bibinfo {author} {\bibfnamefont {M.}~\bibnamefont {Manago}},
  \bibinfo {author} {\bibfnamefont {S.}~\bibnamefont {Kitagawa}}, \bibinfo
  {author} {\bibfnamefont {Z.}~\bibnamefont {Mao}}, \bibinfo {author}
  {\bibfnamefont {S.}~\bibnamefont {Yonezawa}}, \bibinfo {author}
  {\bibfnamefont {Y.}~\bibnamefont {Maeno}}, \ and\ \bibinfo {author}
  {\bibfnamefont {K.}~\bibnamefont {Ishida}},\ }\href@noop {} {\bibfield
  {journal} {\bibinfo  {journal} {Science}\ }\textbf {\bibinfo {volume}
  {376}},\ \bibinfo {pages} {397} (\bibinfo {year} {2022})}\BibitemShut
  {NoStop}%
\bibitem [{\citenamefont {Kitagawa}\ \emph {et~al.}(2018)\citenamefont
  {Kitagawa}, \citenamefont {Nakamine}, \citenamefont {Ishida}, \citenamefont
  {Jeevan}, \citenamefont {Geibel},\ and\ \citenamefont
  {Steglich}}]{PhysRevLett.121.157004}%
  \BibitemOpen
  \bibfield  {author} {\bibinfo {author} {\bibfnamefont {S.}~\bibnamefont
  {Kitagawa}}, \bibinfo {author} {\bibfnamefont {G.}~\bibnamefont {Nakamine}},
  \bibinfo {author} {\bibfnamefont {K.}~\bibnamefont {Ishida}}, \bibinfo
  {author} {\bibfnamefont {H.~S.}\ \bibnamefont {Jeevan}}, \bibinfo {author}
  {\bibfnamefont {C.}~\bibnamefont {Geibel}}, \ and\ \bibinfo {author}
  {\bibfnamefont {F.}~\bibnamefont {Steglich}},\ }\href {\doibase
  10.1103/PhysRevLett.121.157004} {\bibfield  {journal} {\bibinfo  {journal}
  {Phys. Rev. Lett.}\ }\textbf {\bibinfo {volume} {121}},\ \bibinfo {pages}
  {157004} (\bibinfo {year} {2018})}\BibitemShut {NoStop}%
\bibitem [{\citenamefont {Burkhardt}\ and\ \citenamefont
  {Rainer}(1994)}]{burkhart:annphys1994}%
  \BibitemOpen
  \bibfield  {author} {\bibinfo {author} {\bibfnamefont {H.}~\bibnamefont
  {Burkhardt}}\ and\ \bibinfo {author} {\bibfnamefont {D.}~\bibnamefont
  {Rainer}},\ }\href {\doibase https://doi.org/10.1002/andp.19945060305}
  {\bibfield  {journal} {\bibinfo  {journal} {Annalen der Physik}\ }\textbf
  {\bibinfo {volume} {506}},\ \bibinfo {pages} {181} (\bibinfo {year}
  {1994})}\BibitemShut {NoStop}%
\bibitem [{\citenamefont {Matsuo}\ \emph {et~al.}(1998)\citenamefont {Matsuo},
  \citenamefont {Higashitani}, \citenamefont {Nagato},\ and\ \citenamefont
  {Nagai}}]{matsuo:jpsj1998}%
  \BibitemOpen
  \bibfield  {author} {\bibinfo {author} {\bibfnamefont {S.}~\bibnamefont
  {Matsuo}}, \bibinfo {author} {\bibfnamefont {S.}~\bibnamefont {Higashitani}},
  \bibinfo {author} {\bibfnamefont {Y.}~\bibnamefont {Nagato}}, \ and\ \bibinfo
  {author} {\bibfnamefont {K.}~\bibnamefont {Nagai}},\ }\href {\doibase
  10.1143/JPSJ.67.280} {\bibfield  {journal} {\bibinfo  {journal} {Journal of
  the Physical Society of Japan}\ }\textbf {\bibinfo {volume} {67}},\ \bibinfo
  {pages} {280} (\bibinfo {year} {1998})}\BibitemShut {NoStop}%
\bibitem [{\citenamefont {Bulaevskii}\ \emph {et~al.}(1977)\citenamefont
  {Bulaevskii}, \citenamefont {Kuzii},\ and\ \citenamefont
  {Sobyanin}}]{bulaevskii:jetplett1977}%
  \BibitemOpen
  \bibfield  {author} {\bibinfo {author} {\bibfnamefont {L.~N.}\ \bibnamefont
  {Bulaevskii}}, \bibinfo {author} {\bibfnamefont {V.~V.}\ \bibnamefont
  {Kuzii}}, \ and\ \bibinfo {author} {\bibfnamefont {A.~A.}\ \bibnamefont
  {Sobyanin}},\ }\href@noop {} {\bibfield  {journal} {\bibinfo  {journal} {JETP
  Lett.}\ }\textbf {\bibinfo {volume} {25}},\ \bibinfo {pages} {291} (\bibinfo
  {year} {1977})}\BibitemShut {NoStop}%
\bibitem [{\citenamefont {Buzdin}\ \emph {et~al.}(1982)\citenamefont {Buzdin},
  \citenamefont {Bulaevskii},\ and\ \citenamefont
  {Panyuko}}]{buzdin:jetplett1982}%
  \BibitemOpen
  \bibfield  {author} {\bibinfo {author} {\bibfnamefont {A.~I.}\ \bibnamefont
  {Buzdin}}, \bibinfo {author} {\bibfnamefont {L.~N.}\ \bibnamefont
  {Bulaevskii}}, \ and\ \bibinfo {author} {\bibfnamefont {S.~V.}\ \bibnamefont
  {Panyuko}},\ }\href@noop {} {\bibfield  {journal} {\bibinfo  {journal} {JETP
  Lett.}\ }\textbf {\bibinfo {volume} {35}},\ \bibinfo {pages} {179} (\bibinfo
  {year} {1982})}\BibitemShut {NoStop}%
\bibitem [{\citenamefont {Ryazanov}\ \emph {et~al.}(2001)\citenamefont
  {Ryazanov}, \citenamefont {Oboznov}, \citenamefont {Rusanov}, \citenamefont
  {Veretennikov}, \citenamefont {Golubov},\ and\ \citenamefont
  {Aarts}}]{ryazanov:prl2001}%
  \BibitemOpen
  \bibfield  {author} {\bibinfo {author} {\bibfnamefont {V.~V.}\ \bibnamefont
  {Ryazanov}}, \bibinfo {author} {\bibfnamefont {V.~A.}\ \bibnamefont
  {Oboznov}}, \bibinfo {author} {\bibfnamefont {A.~Y.}\ \bibnamefont
  {Rusanov}}, \bibinfo {author} {\bibfnamefont {A.~V.}\ \bibnamefont
  {Veretennikov}}, \bibinfo {author} {\bibfnamefont {A.~A.}\ \bibnamefont
  {Golubov}}, \ and\ \bibinfo {author} {\bibfnamefont {J.}~\bibnamefont
  {Aarts}},\ }\href {\doibase 10.1103/PhysRevLett.86.2427} {\bibfield
  {journal} {\bibinfo  {journal} {Phys. Rev. Lett.}\ }\textbf {\bibinfo
  {volume} {86}},\ \bibinfo {pages} {2427} (\bibinfo {year}
  {2001})}\BibitemShut {NoStop}%
\bibitem [{\citenamefont {Kontos}\ \emph {et~al.}(2002)\citenamefont {Kontos},
  \citenamefont {Aprili}, \citenamefont {Lesueur}, \citenamefont {Gen\^et},
  \citenamefont {Stephanidis},\ and\ \citenamefont
  {Boursier}}]{kontos:prl2002}%
  \BibitemOpen
  \bibfield  {author} {\bibinfo {author} {\bibfnamefont {T.}~\bibnamefont
  {Kontos}}, \bibinfo {author} {\bibfnamefont {M.}~\bibnamefont {Aprili}},
  \bibinfo {author} {\bibfnamefont {J.}~\bibnamefont {Lesueur}}, \bibinfo
  {author} {\bibfnamefont {F.}~\bibnamefont {Gen\^et}}, \bibinfo {author}
  {\bibfnamefont {B.}~\bibnamefont {Stephanidis}}, \ and\ \bibinfo {author}
  {\bibfnamefont {R.}~\bibnamefont {Boursier}},\ }\href {\doibase
  10.1103/PhysRevLett.89.137007} {\bibfield  {journal} {\bibinfo  {journal}
  {Phys. Rev. Lett.}\ }\textbf {\bibinfo {volume} {89}},\ \bibinfo {pages}
  {137007} (\bibinfo {year} {2002})}\BibitemShut {NoStop}%
\bibitem [{\citenamefont {Mironov}\ \emph {et~al.}(2012)\citenamefont
  {Mironov}, \citenamefont {Mel'nikov},\ and\ \citenamefont
  {Buzdin}}]{mironov:prl2012}%
  \BibitemOpen
  \bibfield  {author} {\bibinfo {author} {\bibfnamefont {S.}~\bibnamefont
  {Mironov}}, \bibinfo {author} {\bibfnamefont {A.}~\bibnamefont {Mel'nikov}},
  \ and\ \bibinfo {author} {\bibfnamefont {A.}~\bibnamefont {Buzdin}},\ }\href
  {\doibase 10.1103/PhysRevLett.109.237002} {\bibfield  {journal} {\bibinfo
  {journal} {Phys. Rev. Lett.}\ }\textbf {\bibinfo {volume} {109}},\ \bibinfo
  {pages} {237002} (\bibinfo {year} {2012})}\BibitemShut {NoStop}%
\bibitem [{\citenamefont {M{\'e}nard}\ \emph {et~al.}(2017)\citenamefont
  {M{\'e}nard}, \citenamefont {Guissart}, \citenamefont {Brun}, \citenamefont
  {Leriche}, \citenamefont {Trif}, \citenamefont {Debontridder}, \citenamefont
  {Demaille}, \citenamefont {Roditchev}, \citenamefont {Simon},\ and\
  \citenamefont {Cren}}]{cren:naturecommun2017}%
  \BibitemOpen
  \bibfield  {author} {\bibinfo {author} {\bibfnamefont {G.~C.}\ \bibnamefont
  {M{\'e}nard}}, \bibinfo {author} {\bibfnamefont {S.}~\bibnamefont
  {Guissart}}, \bibinfo {author} {\bibfnamefont {C.}~\bibnamefont {Brun}},
  \bibinfo {author} {\bibfnamefont {R.~T.}\ \bibnamefont {Leriche}}, \bibinfo
  {author} {\bibfnamefont {M.}~\bibnamefont {Trif}}, \bibinfo {author}
  {\bibfnamefont {F.}~\bibnamefont {Debontridder}}, \bibinfo {author}
  {\bibfnamefont {D.}~\bibnamefont {Demaille}}, \bibinfo {author}
  {\bibfnamefont {D.}~\bibnamefont {Roditchev}}, \bibinfo {author}
  {\bibfnamefont {P.}~\bibnamefont {Simon}}, \ and\ \bibinfo {author}
  {\bibfnamefont {T.}~\bibnamefont {Cren}},\ }\href {\doibase
  10.1038/s41467-017-02192-x} {\bibfield  {journal} {\bibinfo  {journal} {Nat.
  Commun.}\ }\textbf {\bibinfo {volume} {8}},\ \bibinfo {pages} {2040}
  (\bibinfo {year} {2017})}\BibitemShut {NoStop}%
\bibitem [{\citenamefont {M{\'e}nard}\ \emph
  {et~al.}(2019{\natexlab{a}})\citenamefont {M{\'e}nard}, \citenamefont {Brun},
  \citenamefont {Leriche}, \citenamefont {Trif}, \citenamefont {Debontridder},
  \citenamefont {Demaille}, \citenamefont {Roditchev}, \citenamefont {Simon},\
  and\ \citenamefont {Cren}}]{menard2019yu}%
  \BibitemOpen
  \bibfield  {author} {\bibinfo {author} {\bibfnamefont {G.~C.}\ \bibnamefont
  {M{\'e}nard}}, \bibinfo {author} {\bibfnamefont {C.}~\bibnamefont {Brun}},
  \bibinfo {author} {\bibfnamefont {R.}~\bibnamefont {Leriche}}, \bibinfo
  {author} {\bibfnamefont {M.}~\bibnamefont {Trif}}, \bibinfo {author}
  {\bibfnamefont {F.}~\bibnamefont {Debontridder}}, \bibinfo {author}
  {\bibfnamefont {D.}~\bibnamefont {Demaille}}, \bibinfo {author}
  {\bibfnamefont {D.}~\bibnamefont {Roditchev}}, \bibinfo {author}
  {\bibfnamefont {P.}~\bibnamefont {Simon}}, \ and\ \bibinfo {author}
  {\bibfnamefont {T.}~\bibnamefont {Cren}},\ }\href {\doibase
  10.1140/epjst/e2018-800056-3} {\bibfield  {journal} {\bibinfo  {journal}
  {Eur. Phys. J. Special Topics}\ }\textbf {\bibinfo {volume} {227}},\ \bibinfo
  {pages} {2303} (\bibinfo {year} {2019}{\natexlab{a}})}\BibitemShut {NoStop}%
\bibitem [{\citenamefont {M{\'e}nard}\ \emph
  {et~al.}(2019{\natexlab{b}})\citenamefont {M{\'e}nard}, \citenamefont
  {Mesaros}, \citenamefont {Brun}, \citenamefont {Debontridder}, \citenamefont
  {Roditchev}, \citenamefont {Simon},\ and\ \citenamefont
  {Cren}}]{menard2019isolated}%
  \BibitemOpen
  \bibfield  {author} {\bibinfo {author} {\bibfnamefont {G.~C.}\ \bibnamefont
  {M{\'e}nard}}, \bibinfo {author} {\bibfnamefont {A.}~\bibnamefont {Mesaros}},
  \bibinfo {author} {\bibfnamefont {C.}~\bibnamefont {Brun}}, \bibinfo {author}
  {\bibfnamefont {F.}~\bibnamefont {Debontridder}}, \bibinfo {author}
  {\bibfnamefont {D.}~\bibnamefont {Roditchev}}, \bibinfo {author}
  {\bibfnamefont {P.}~\bibnamefont {Simon}}, \ and\ \bibinfo {author}
  {\bibfnamefont {T.}~\bibnamefont {Cren}},\ }\href {\doibase
  10.1038/s41467-019-10397-5} {\bibfield  {journal} {\bibinfo  {journal} {Nat.
  Commun.}\ }\textbf {\bibinfo {volume} {10}},\ \bibinfo {pages} {2587}
  (\bibinfo {year} {2019}{\natexlab{b}})}\BibitemShut {NoStop}%
\bibitem [{\citenamefont {Abrikosov}\ and\ \citenamefont
  {Gor{\textquoteright}kov}(1961)}]{abrikosov:jetp1961}%
  \BibitemOpen
  \bibfield  {author} {\bibinfo {author} {\bibfnamefont {A.~A.}\ \bibnamefont
  {Abrikosov}}\ and\ \bibinfo {author} {\bibfnamefont {L.~P.}\ \bibnamefont
  {Gor{\textquoteright}kov}},\ }\href@noop {} {\bibfield  {journal} {\bibinfo
  {journal} {Sov. Phys. JETP}\ }\textbf {\bibinfo {volume} {12}},\ \bibinfo
  {pages} {1243} (\bibinfo {year} {1961})}\BibitemShut {NoStop}%
\bibitem [{\citenamefont {Yu}(1965)}]{yu:actphys1965}%
  \BibitemOpen
  \bibfield  {author} {\bibinfo {author} {\bibfnamefont {L.}~\bibnamefont
  {Yu}},\ }\href@noop {} {\bibfield  {journal} {\bibinfo  {journal} {Acta.
  Phys. Sin}\ }\textbf {\bibinfo {volume} {21}},\ \bibinfo {pages} {75}
  (\bibinfo {year} {1965})}\BibitemShut {NoStop}%
\bibitem [{\citenamefont {Shiba}(1968)}]{shiba:ptp1968}%
  \BibitemOpen
  \bibfield  {author} {\bibinfo {author} {\bibfnamefont {H.}~\bibnamefont
  {Shiba}},\ }\href {\doibase 10.1143/PTP.40.435} {\bibfield  {journal}
  {\bibinfo  {journal} {Progress of Theoretical Physics}\ }\textbf {\bibinfo
  {volume} {40}},\ \bibinfo {pages} {435} (\bibinfo {year} {1968})}\BibitemShut
  {NoStop}%
\bibitem [{\citenamefont {Rusinov}(1969{\natexlab{a}})}]{rusinov:jetplett1969}%
  \BibitemOpen
  \bibfield  {author} {\bibinfo {author} {\bibfnamefont {A.~I.}\ \bibnamefont
  {Rusinov}},\ }\href@noop {} {\bibfield  {journal} {\bibinfo  {journal} {Sov.
  Phys. JETP Lett}\ }\textbf {\bibinfo {volume} {9}},\ \bibinfo {pages} {85}
  (\bibinfo {year} {1969}{\natexlab{a}})}\BibitemShut {NoStop}%
\bibitem [{\citenamefont {Yazdani}\ \emph {et~al.}(1997)\citenamefont
  {Yazdani}, \citenamefont {Jones}, \citenamefont {Lutz}, \citenamefont
  {Crommie},\ and\ \citenamefont {Eigler}}]{yazdani:science1997}%
  \BibitemOpen
  \bibfield  {author} {\bibinfo {author} {\bibfnamefont {A.}~\bibnamefont
  {Yazdani}}, \bibinfo {author} {\bibfnamefont {B.~A.}\ \bibnamefont {Jones}},
  \bibinfo {author} {\bibfnamefont {C.~P.}\ \bibnamefont {Lutz}}, \bibinfo
  {author} {\bibfnamefont {M.~F.}\ \bibnamefont {Crommie}}, \ and\ \bibinfo
  {author} {\bibfnamefont {D.~M.}\ \bibnamefont {Eigler}},\ }\href {\doibase
  10.1126/science.275.5307.1767} {\bibfield  {journal} {\bibinfo  {journal}
  {Science}\ }\textbf {\bibinfo {volume} {275}},\ \bibinfo {pages} {1767}
  (\bibinfo {year} {1997})}\BibitemShut {NoStop}%
\bibitem [{\citenamefont {Choy}\ \emph {et~al.}(2011)\citenamefont {Choy},
  \citenamefont {Edge}, \citenamefont {Akhmerov},\ and\ \citenamefont
  {Beenakker}}]{choy:prb2011}%
  \BibitemOpen
  \bibfield  {author} {\bibinfo {author} {\bibfnamefont {T.-P.}\ \bibnamefont
  {Choy}}, \bibinfo {author} {\bibfnamefont {J.~M.}\ \bibnamefont {Edge}},
  \bibinfo {author} {\bibfnamefont {A.~R.}\ \bibnamefont {Akhmerov}}, \ and\
  \bibinfo {author} {\bibfnamefont {C.~W.~J.}\ \bibnamefont {Beenakker}},\
  }\href {\doibase 10.1103/PhysRevB.84.195442} {\bibfield  {journal} {\bibinfo
  {journal} {Phys. Rev. B}\ }\textbf {\bibinfo {volume} {84}},\ \bibinfo
  {pages} {195442} (\bibinfo {year} {2011})}\BibitemShut {NoStop}%
\bibitem [{\citenamefont {Nadj-Perge}\ \emph {et~al.}(2014)\citenamefont
  {Nadj-Perge}, \citenamefont {Drozdov}, \citenamefont {Li}, \citenamefont
  {Chen}, \citenamefont {Jeon}, \citenamefont {Seo}, \citenamefont {MacDonald},
  \citenamefont {Bernevig},\ and\ \citenamefont
  {Yazdani}}]{Nadjerge:science2014}%
  \BibitemOpen
  \bibfield  {author} {\bibinfo {author} {\bibfnamefont {S.}~\bibnamefont
  {Nadj-Perge}}, \bibinfo {author} {\bibfnamefont {I.~K.}\ \bibnamefont
  {Drozdov}}, \bibinfo {author} {\bibfnamefont {J.}~\bibnamefont {Li}},
  \bibinfo {author} {\bibfnamefont {H.}~\bibnamefont {Chen}}, \bibinfo {author}
  {\bibfnamefont {S.}~\bibnamefont {Jeon}}, \bibinfo {author} {\bibfnamefont
  {J.}~\bibnamefont {Seo}}, \bibinfo {author} {\bibfnamefont {A.~H.}\
  \bibnamefont {MacDonald}}, \bibinfo {author} {\bibfnamefont {B.~A.}\
  \bibnamefont {Bernevig}}, \ and\ \bibinfo {author} {\bibfnamefont
  {A.}~\bibnamefont {Yazdani}},\ }\href {\doibase 10.1126/science.1259327}
  {\bibfield  {journal} {\bibinfo  {journal} {Science}\ }\textbf {\bibinfo
  {volume} {346}},\ \bibinfo {pages} {602} (\bibinfo {year}
  {2014})}\BibitemShut {NoStop}%
\bibitem [{\citenamefont {Rusinov}(1969{\natexlab{b}})}]{rusinov:jetp1969}%
  \BibitemOpen
  \bibfield  {author} {\bibinfo {author} {\bibfnamefont {A.~I.}\ \bibnamefont
  {Rusinov}},\ }\href@noop {} {\bibfield  {journal} {\bibinfo  {journal} {Sov.
  Phys. JETP}\ }\textbf {\bibinfo {volume} {29}},\ \bibinfo {pages} {1101}
  (\bibinfo {year} {1969}{\natexlab{b}})}\BibitemShut {NoStop}%
\bibitem [{\citenamefont {Fominov}\ and\ \citenamefont
  {Skvortsov}(2016)}]{fominov:prb2016}%
  \BibitemOpen
  \bibfield  {author} {\bibinfo {author} {\bibfnamefont {Y.~V.}\ \bibnamefont
  {Fominov}}\ and\ \bibinfo {author} {\bibfnamefont {M.~A.}\ \bibnamefont
  {Skvortsov}},\ }\href {\doibase 10.1103/PhysRevB.93.144511} {\bibfield
  {journal} {\bibinfo  {journal} {Phys. Rev. B}\ }\textbf {\bibinfo {volume}
  {93}},\ \bibinfo {pages} {144511} (\bibinfo {year} {2016})}\BibitemShut
  {NoStop}%
\bibitem [{\citenamefont {Salkola}\ \emph {et~al.}(1997)\citenamefont
  {Salkola}, \citenamefont {Balatsky},\ and\ \citenamefont
  {Schrieffer}}]{salkola:prb1997}%
  \BibitemOpen
  \bibfield  {author} {\bibinfo {author} {\bibfnamefont {M.~I.}\ \bibnamefont
  {Salkola}}, \bibinfo {author} {\bibfnamefont {A.~V.}\ \bibnamefont
  {Balatsky}}, \ and\ \bibinfo {author} {\bibfnamefont {J.~R.}\ \bibnamefont
  {Schrieffer}},\ }\href {\doibase 10.1103/PhysRevB.55.12648} {\bibfield
  {journal} {\bibinfo  {journal} {Phys. Rev. B}\ }\textbf {\bibinfo {volume}
  {55}},\ \bibinfo {pages} {12648} (\bibinfo {year} {1997})}\BibitemShut
  {NoStop}%
\bibitem [{\citenamefont {Flatt\'e}\ and\ \citenamefont
  {Byers}(1997)}]{flatte:prl1997}%
  \BibitemOpen
  \bibfield  {author} {\bibinfo {author} {\bibfnamefont {M.~E.}\ \bibnamefont
  {Flatt\'e}}\ and\ \bibinfo {author} {\bibfnamefont {J.~M.}\ \bibnamefont
  {Byers}},\ }\href {\doibase 10.1103/PhysRevLett.78.3761} {\bibfield
  {journal} {\bibinfo  {journal} {Phys. Rev. Lett.}\ }\textbf {\bibinfo
  {volume} {78}},\ \bibinfo {pages} {3761} (\bibinfo {year}
  {1997})}\BibitemShut {NoStop}%
\bibitem [{\citenamefont {Balatsky}\ \emph {et~al.}(2006)\citenamefont
  {Balatsky}, \citenamefont {Vekhter},\ and\ \citenamefont
  {Zhu}}]{balatsky:rmp2006}%
  \BibitemOpen
  \bibfield  {author} {\bibinfo {author} {\bibfnamefont {A.~V.}\ \bibnamefont
  {Balatsky}}, \bibinfo {author} {\bibfnamefont {I.}~\bibnamefont {Vekhter}}, \
  and\ \bibinfo {author} {\bibfnamefont {J.-X.}\ \bibnamefont {Zhu}},\ }\href
  {\doibase 10.1103/RevModPhys.78.373} {\bibfield  {journal} {\bibinfo
  {journal} {Rev. Mod. Phys.}\ }\textbf {\bibinfo {volume} {78}},\ \bibinfo
  {pages} {373} (\bibinfo {year} {2006})}\BibitemShut {NoStop}%
\bibitem [{\citenamefont {Suzuki}\ \emph {et~al.}(2022)\citenamefont {Suzuki},
  \citenamefont {Sato},\ and\ \citenamefont {Asano}}]{shu:prb2022}%
  \BibitemOpen
  \bibfield  {author} {\bibinfo {author} {\bibfnamefont {S.-I.}\ \bibnamefont
  {Suzuki}}, \bibinfo {author} {\bibfnamefont {T.}~\bibnamefont {Sato}}, \ and\
  \bibinfo {author} {\bibfnamefont {Y.}~\bibnamefont {Asano}},\ }\href
  {\doibase 10.1103/PhysRevB.106.104518} {\bibfield  {journal} {\bibinfo
  {journal} {Phys. Rev. B}\ }\textbf {\bibinfo {volume} {106}},\ \bibinfo
  {pages} {104518} (\bibinfo {year} {2022})}\BibitemShut {NoStop}%
\bibitem [{\citenamefont {Kuzmanovski}\ \emph {et~al.}(2020)\citenamefont
  {Kuzmanovski}, \citenamefont {Souto},\ and\ \citenamefont
  {Balatsky}}]{kuzmanovski:prb2020}%
  \BibitemOpen
  \bibfield  {author} {\bibinfo {author} {\bibfnamefont {D.}~\bibnamefont
  {Kuzmanovski}}, \bibinfo {author} {\bibfnamefont {R.~S.}\ \bibnamefont
  {Souto}}, \ and\ \bibinfo {author} {\bibfnamefont {A.~V.}\ \bibnamefont
  {Balatsky}},\ }\href {\doibase 10.1103/PhysRevB.101.094505} {\bibfield
  {journal} {\bibinfo  {journal} {Phys. Rev. B}\ }\textbf {\bibinfo {volume}
  {101}},\ \bibinfo {pages} {094505} (\bibinfo {year} {2020})}\BibitemShut
  {NoStop}%
\bibitem [{\citenamefont {Perrin}\ \emph {et~al.}(2020)\citenamefont {Perrin},
  \citenamefont {Santos}, \citenamefont {M\'enard}, \citenamefont {Brun},
  \citenamefont {Cren}, \citenamefont {Civelli},\ and\ \citenamefont
  {Simon}}]{perrin:prl2020}%
  \BibitemOpen
  \bibfield  {author} {\bibinfo {author} {\bibfnamefont {V.}~\bibnamefont
  {Perrin}}, \bibinfo {author} {\bibfnamefont {F.~L.~N.}\ \bibnamefont
  {Santos}}, \bibinfo {author} {\bibfnamefont {G.~C.}\ \bibnamefont
  {M\'enard}}, \bibinfo {author} {\bibfnamefont {C.}~\bibnamefont {Brun}},
  \bibinfo {author} {\bibfnamefont {T.}~\bibnamefont {Cren}}, \bibinfo {author}
  {\bibfnamefont {M.}~\bibnamefont {Civelli}}, \ and\ \bibinfo {author}
  {\bibfnamefont {P.}~\bibnamefont {Simon}},\ }\href {\doibase
  10.1103/PhysRevLett.125.117003} {\bibfield  {journal} {\bibinfo  {journal}
  {Phys. Rev. Lett.}\ }\textbf {\bibinfo {volume} {125}},\ \bibinfo {pages}
  {117003} (\bibinfo {year} {2020})}\BibitemShut {NoStop}%
\bibitem [{\citenamefont {Bergeret}\ \emph {et~al.}(2001)\citenamefont
  {Bergeret}, \citenamefont {Volkov},\ and\ \citenamefont
  {Efetov}}]{bergeret:prl2001}%
  \BibitemOpen
  \bibfield  {author} {\bibinfo {author} {\bibfnamefont {F.~S.}\ \bibnamefont
  {Bergeret}}, \bibinfo {author} {\bibfnamefont {A.~F.}\ \bibnamefont
  {Volkov}}, \ and\ \bibinfo {author} {\bibfnamefont {K.~B.}\ \bibnamefont
  {Efetov}},\ }\href {\doibase 10.1103/PhysRevLett.86.4096} {\bibfield
  {journal} {\bibinfo  {journal} {Phys. Rev. Lett.}\ }\textbf {\bibinfo
  {volume} {86}},\ \bibinfo {pages} {4096} (\bibinfo {year}
  {2001})}\BibitemShut {NoStop}%
\bibitem [{\citenamefont {Keizer}\ \emph {et~al.}(2006)\citenamefont {Keizer},
  \citenamefont {Goennenwein}, \citenamefont {Klapwijk}, \citenamefont {Miao},
  \citenamefont {Xiao},\ and\ \citenamefont {Gupta}}]{keizer:nature2006}%
  \BibitemOpen
  \bibfield  {author} {\bibinfo {author} {\bibfnamefont {R.~S.}\ \bibnamefont
  {Keizer}}, \bibinfo {author} {\bibfnamefont {S.~T.~B.}\ \bibnamefont
  {Goennenwein}}, \bibinfo {author} {\bibfnamefont {T.~M.}\ \bibnamefont
  {Klapwijk}}, \bibinfo {author} {\bibfnamefont {G.}~\bibnamefont {Miao}},
  \bibinfo {author} {\bibfnamefont {G.}~\bibnamefont {Xiao}}, \ and\ \bibinfo
  {author} {\bibfnamefont {A.}~\bibnamefont {Gupta}},\ }\href {\doibase
  10.1038/nature04499} {\bibfield  {journal} {\bibinfo  {journal} {Nature}\
  }\textbf {\bibinfo {volume} {439}},\ \bibinfo {pages} {825} (\bibinfo {year}
  {2006})}\BibitemShut {NoStop}%
\bibitem [{\citenamefont {Asano}\ \emph
  {et~al.}(2007{\natexlab{a}})\citenamefont {Asano}, \citenamefont {Tanaka},\
  and\ \citenamefont {Golubov}}]{asano:prl2007sfs}%
  \BibitemOpen
  \bibfield  {author} {\bibinfo {author} {\bibfnamefont {Y.}~\bibnamefont
  {Asano}}, \bibinfo {author} {\bibfnamefont {Y.}~\bibnamefont {Tanaka}}, \
  and\ \bibinfo {author} {\bibfnamefont {A.~A.}\ \bibnamefont {Golubov}},\
  }\href {\doibase 10.1103/PhysRevLett.98.107002} {\bibfield  {journal}
  {\bibinfo  {journal} {Phys. Rev. Lett.}\ }\textbf {\bibinfo {volume} {98}},\
  \bibinfo {pages} {107002} (\bibinfo {year} {2007}{\natexlab{a}})}\BibitemShut
  {NoStop}%
\bibitem [{\citenamefont {Asano}\ \emph
  {et~al.}(2007{\natexlab{b}})\citenamefont {Asano}, \citenamefont {Sawa},
  \citenamefont {Tanaka},\ and\ \citenamefont {Golubov}}]{asano:prb2007sfs}%
  \BibitemOpen
  \bibfield  {author} {\bibinfo {author} {\bibfnamefont {Y.}~\bibnamefont
  {Asano}}, \bibinfo {author} {\bibfnamefont {Y.}~\bibnamefont {Sawa}},
  \bibinfo {author} {\bibfnamefont {Y.}~\bibnamefont {Tanaka}}, \ and\ \bibinfo
  {author} {\bibfnamefont {A.~A.}\ \bibnamefont {Golubov}},\ }\href {\doibase
  10.1103/PhysRevB.76.224525} {\bibfield  {journal} {\bibinfo  {journal} {Phys.
  Rev. B}\ }\textbf {\bibinfo {volume} {76}},\ \bibinfo {pages} {224525}
  (\bibinfo {year} {2007}{\natexlab{b}})}\BibitemShut {NoStop}%
\bibitem [{\citenamefont {Braude}\ and\ \citenamefont
  {Nazarov}(2007)}]{braude:prl2007}%
  \BibitemOpen
  \bibfield  {author} {\bibinfo {author} {\bibfnamefont {V.}~\bibnamefont
  {Braude}}\ and\ \bibinfo {author} {\bibfnamefont {Y.~V.}\ \bibnamefont
  {Nazarov}},\ }\href {\doibase 10.1103/PhysRevLett.98.077003} {\bibfield
  {journal} {\bibinfo  {journal} {Phys. Rev. Lett.}\ }\textbf {\bibinfo
  {volume} {98}},\ \bibinfo {pages} {077003} (\bibinfo {year}
  {2007})}\BibitemShut {NoStop}%
\bibitem [{\citenamefont {Robinson}\ \emph {et~al.}(2010)\citenamefont
  {Robinson}, \citenamefont {Witt},\ and\ \citenamefont
  {Blamire}}]{robinson:science2010}%
  \BibitemOpen
  \bibfield  {author} {\bibinfo {author} {\bibfnamefont {J.~W.~A.}\
  \bibnamefont {Robinson}}, \bibinfo {author} {\bibfnamefont {J.~D.~S.}\
  \bibnamefont {Witt}}, \ and\ \bibinfo {author} {\bibfnamefont {M.~G.}\
  \bibnamefont {Blamire}},\ }\href {\doibase 10.1126/science.1189246}
  {\bibfield  {journal} {\bibinfo  {journal} {Science}\ }\textbf {\bibinfo
  {volume} {329}},\ \bibinfo {pages} {59} (\bibinfo {year} {2010})}\BibitemShut
  {NoStop}%
\bibitem [{\citenamefont {Khaire}\ \emph {et~al.}(2010)\citenamefont {Khaire},
  \citenamefont {Khasawneh}, \citenamefont {Pratt},\ and\ \citenamefont
  {Birge}}]{birge:prl2010}%
  \BibitemOpen
  \bibfield  {author} {\bibinfo {author} {\bibfnamefont {T.~S.}\ \bibnamefont
  {Khaire}}, \bibinfo {author} {\bibfnamefont {M.~A.}\ \bibnamefont
  {Khasawneh}}, \bibinfo {author} {\bibfnamefont {W.~P.}\ \bibnamefont
  {Pratt}}, \ and\ \bibinfo {author} {\bibfnamefont {N.~O.}\ \bibnamefont
  {Birge}},\ }\href {\doibase 10.1103/PhysRevLett.104.137002} {\bibfield
  {journal} {\bibinfo  {journal} {Phys. Rev. Lett.}\ }\textbf {\bibinfo
  {volume} {104}},\ \bibinfo {pages} {137002} (\bibinfo {year}
  {2010})}\BibitemShut {NoStop}%
\bibitem [{\citenamefont {Anwar}\ \emph {et~al.}(2010)\citenamefont {Anwar},
  \citenamefont {Czeschka}, \citenamefont {Hesselberth}, \citenamefont
  {Porcu},\ and\ \citenamefont {Aarts}}]{anwar:prb2010}%
  \BibitemOpen
  \bibfield  {author} {\bibinfo {author} {\bibfnamefont {M.~S.}\ \bibnamefont
  {Anwar}}, \bibinfo {author} {\bibfnamefont {F.}~\bibnamefont {Czeschka}},
  \bibinfo {author} {\bibfnamefont {M.}~\bibnamefont {Hesselberth}}, \bibinfo
  {author} {\bibfnamefont {M.}~\bibnamefont {Porcu}}, \ and\ \bibinfo {author}
  {\bibfnamefont {J.}~\bibnamefont {Aarts}},\ }\href {\doibase
  10.1103/PhysRevB.82.100501} {\bibfield  {journal} {\bibinfo  {journal} {Phys.
  Rev. B}\ }\textbf {\bibinfo {volume} {82}},\ \bibinfo {pages} {100501}
  (\bibinfo {year} {2010})}\BibitemShut {NoStop}%
\bibitem [{\citenamefont {Eilenberger}(1968)}]{eilenberger:zphys1968}%
  \BibitemOpen
  \bibfield  {author} {\bibinfo {author} {\bibfnamefont {G.}~\bibnamefont
  {Eilenberger}},\ }\href {\doibase 10.1007/BF01379803} {\bibfield  {journal}
  {\bibinfo  {journal} {Zeitschrift f{\"u}r Physik A Hadrons and nuclei}\
  }\textbf {\bibinfo {volume} {214}},\ \bibinfo {pages} {195} (\bibinfo {year}
  {1968})}\BibitemShut {NoStop}%
\bibitem [{\citenamefont {Schopohl}(1998)}]{schopohl:arxiv1998}%
  \BibitemOpen
  \bibfield  {author} {\bibinfo {author} {\bibfnamefont {N.}~\bibnamefont
  {Schopohl}},\ }\href@noop {} {\bibfield  {journal} {\bibinfo  {journal}
  {arXiv:cond-mat}\ ,\ \bibinfo {pages} {9804064}} (\bibinfo {year}
  {1998})}\BibitemShut {NoStop}%
\bibitem [{\citenamefont {Schopohl}\ and\ \citenamefont
  {Maki}(1995)}]{Schopohl_PRB_95}%
  \BibitemOpen
  \bibfield  {author} {\bibinfo {author} {\bibfnamefont {N.}~\bibnamefont
  {Schopohl}}\ and\ \bibinfo {author} {\bibfnamefont {K.}~\bibnamefont
  {Maki}},\ }\href {\doibase 10.1103/PhysRevB.52.490} {\bibfield  {journal}
  {\bibinfo  {journal} {Phys. Rev. B}\ }\textbf {\bibinfo {volume} {52}},\
  \bibinfo {pages} {490} (\bibinfo {year} {1995})}\BibitemShut {NoStop}%
\bibitem [{\citenamefont {Eschrig}(2000)}]{Eschrig_PRB_00}%
  \BibitemOpen
  \bibfield  {author} {\bibinfo {author} {\bibfnamefont {M.}~\bibnamefont
  {Eschrig}},\ }\href {\doibase 10.1103/PhysRevB.61.9061} {\bibfield  {journal}
  {\bibinfo  {journal} {Phys. Rev. B}\ }\textbf {\bibinfo {volume} {61}},\
  \bibinfo {pages} {9061} (\bibinfo {year} {2000})}\BibitemShut {NoStop}%
\bibitem [{\citenamefont {Eschrig}(2009)}]{Eschrig_PRB_09}%
  \BibitemOpen
  \bibfield  {author} {\bibinfo {author} {\bibfnamefont {M.}~\bibnamefont
  {Eschrig}},\ }\href {\doibase 10.1103/PhysRevB.80.134511} {\bibfield
  {journal} {\bibinfo  {journal} {Phys. Rev. B}\ }\textbf {\bibinfo {volume}
  {80}},\ \bibinfo {pages} {134511} (\bibinfo {year} {2009})}\BibitemShut
  {NoStop}%
\bibitem [{Note1()}]{Note1}%
  \BibitemOpen
  \bibinfo {note} {In our configuration, a vortex state could be a possible
  solution of the Eilenberger equation. However, the vortex has typically a
  higher energy than the homogeneous state. Therefore, in this paper, we focus
  on the homogeneous superconductivity at $\rho \gg R_0$.}\BibitemShut {Stop}%
\bibitem [{Note2()}]{Note2}%
  \BibitemOpen
  \bibinfo {note} {{ Inversion symmetry is broken also by the spatially
  oscillating pair potential in the FFLO states. When the inversion symmetry is
  broken, parity is no longer a well-defined symmetry index. In other words,
  the parity mixing among even- and odd-parity pairing functions is allowed.
  Therefore, odd-parity pairs exist in the FFLO state of a spin-singlet
  even-parity superconductor.}}\BibitemShut {Stop}%
\bibitem [{\citenamefont {Eilenberger}(1966)}]{eilenberger:zphys1966}%
  \BibitemOpen
  \bibfield  {author} {\bibinfo {author} {\bibfnamefont {G.}~\bibnamefont
  {Eilenberger}},\ }\href {\doibase 10.1007/BF01327140} {\bibfield  {journal}
  {\bibinfo  {journal} {Zeitschrift f{\"u}r Physik}\ }\textbf {\bibinfo
  {volume} {190}},\ \bibinfo {pages} {142} (\bibinfo {year}
  {1966})}\BibitemShut {NoStop}%
\bibitem [{\citenamefont {Suzuki}\ and\ \citenamefont
  {Asano}(2015)}]{suzuki:prb2015}%
  \BibitemOpen
  \bibfield  {author} {\bibinfo {author} {\bibfnamefont {S.-I.}\ \bibnamefont
  {Suzuki}}\ and\ \bibinfo {author} {\bibfnamefont {Y.}~\bibnamefont {Asano}},\
  }\href {\doibase 10.1103/PhysRevB.91.214510} {\bibfield  {journal} {\bibinfo
  {journal} {Phys. Rev. B}\ }\textbf {\bibinfo {volume} {91}},\ \bibinfo
  {pages} {214510} (\bibinfo {year} {2015})}\BibitemShut {NoStop}%
\bibitem [{\citenamefont {Chandrasekhar}(1962)}]{chandrasekhar:apl1962}%
  \BibitemOpen
  \bibfield  {author} {\bibinfo {author} {\bibfnamefont {B.~S.}\ \bibnamefont
  {Chandrasekhar}},\ }\href {\doibase 10.1063/1.1777362} {\bibfield  {journal}
  {\bibinfo  {journal} {Applied Physics Letters}\ }\textbf {\bibinfo {volume}
  {1}},\ \bibinfo {pages} {7} (\bibinfo {year} {1962})}\BibitemShut {NoStop}%
\bibitem [{\citenamefont {Clogston}(1962)}]{clogston:prl1962}%
  \BibitemOpen
  \bibfield  {author} {\bibinfo {author} {\bibfnamefont {A.~M.}\ \bibnamefont
  {Clogston}},\ }\href {\doibase 10.1103/PhysRevLett.9.266} {\bibfield
  {journal} {\bibinfo  {journal} {Phys. Rev. Lett.}\ }\textbf {\bibinfo
  {volume} {9}},\ \bibinfo {pages} {266} (\bibinfo {year} {1962})}\BibitemShut
  {NoStop}%
\bibitem [{\citenamefont {Asano}\ \emph {et~al.}(2011)\citenamefont {Asano},
  \citenamefont {Golubov}, \citenamefont {Fominov},\ and\ \citenamefont
  {Tanaka}}]{asano:prl2011}%
  \BibitemOpen
  \bibfield  {author} {\bibinfo {author} {\bibfnamefont {Y.}~\bibnamefont
  {Asano}}, \bibinfo {author} {\bibfnamefont {A.~A.}\ \bibnamefont {Golubov}},
  \bibinfo {author} {\bibfnamefont {Y.~V.}\ \bibnamefont {Fominov}}, \ and\
  \bibinfo {author} {\bibfnamefont {Y.}~\bibnamefont {Tanaka}},\ }\href
  {\doibase 10.1103/PhysRevLett.107.087001} {\bibfield  {journal} {\bibinfo
  {journal} {Phys. Rev. Lett.}\ }\textbf {\bibinfo {volume} {107}},\ \bibinfo
  {pages} {087001} (\bibinfo {year} {2011})}\BibitemShut {NoStop}%
\bibitem [{\citenamefont {Asano}\ and\ \citenamefont
  {Sasaki}(2015)}]{asano:prb2015}%
  \BibitemOpen
  \bibfield  {author} {\bibinfo {author} {\bibfnamefont {Y.}~\bibnamefont
  {Asano}}\ and\ \bibinfo {author} {\bibfnamefont {A.}~\bibnamefont {Sasaki}},\
  }\href {\doibase 10.1103/PhysRevB.92.224508} {\bibfield  {journal} {\bibinfo
  {journal} {Phys. Rev. B}\ }\textbf {\bibinfo {volume} {92}},\ \bibinfo
  {pages} {224508} (\bibinfo {year} {2015})}\BibitemShut {NoStop}%
\bibitem [{\citenamefont {Tanaka}\ and\ \citenamefont
  {Golubov}(2007)}]{tanaka:prl2007}%
  \BibitemOpen
  \bibfield  {author} {\bibinfo {author} {\bibfnamefont {Y.}~\bibnamefont
  {Tanaka}}\ and\ \bibinfo {author} {\bibfnamefont {A.~A.}\ \bibnamefont
  {Golubov}},\ }\href {\doibase 10.1103/PhysRevLett.98.037003} {\bibfield
  {journal} {\bibinfo  {journal} {Phys. Rev. Lett.}\ }\textbf {\bibinfo
  {volume} {98}},\ \bibinfo {pages} {037003} (\bibinfo {year}
  {2007})}\BibitemShut {NoStop}%
\bibitem [{\citenamefont {Kim}\ \emph {et~al.}(2021)\citenamefont {Kim},
  \citenamefont {Kobayashi},\ and\ \citenamefont {Asano}}]{kim:jpsj2021}%
  \BibitemOpen
  \bibfield  {author} {\bibinfo {author} {\bibfnamefont {D.}~\bibnamefont
  {Kim}}, \bibinfo {author} {\bibfnamefont {S.}~\bibnamefont {Kobayashi}}, \
  and\ \bibinfo {author} {\bibfnamefont {Y.}~\bibnamefont {Asano}},\ }\href
  {\doibase 10.7566/JPSJ.90.104708} {\bibfield  {journal} {\bibinfo  {journal}
  {J. Phys. Soc. Jpn.}\ }\textbf {\bibinfo {volume} {90}},\ \bibinfo {pages}
  {104708} (\bibinfo {year} {2021})}\BibitemShut {NoStop}%
\bibitem [{\citenamefont {Aslamazov}(1969)}]{aslamazov1969influence}%
  \BibitemOpen
  \bibfield  {author} {\bibinfo {author} {\bibfnamefont {L.~G.}\ \bibnamefont
  {Aslamazov}},\ }\href@noop {} {\bibfield  {journal} {\bibinfo  {journal}
  {Sov. Phys. JETP}\ }\textbf {\bibinfo {volume} {28}},\ \bibinfo {pages} {773}
  (\bibinfo {year} {1969})}\BibitemShut {NoStop}%
\bibitem [{\citenamefont {Takada}(1970)}]{takada1970superconductivity}%
  \BibitemOpen
  \bibfield  {author} {\bibinfo {author} {\bibfnamefont {S.}~\bibnamefont
  {Takada}},\ }\href@noop {} {\bibfield  {journal} {\bibinfo  {journal}
  {Progress of Theoretical Physics}\ }\textbf {\bibinfo {volume} {43}},\
  \bibinfo {pages} {27} (\bibinfo {year} {1970})}\BibitemShut {NoStop}%
\bibitem [{\citenamefont {Houzet}\ and\ \citenamefont
  {Mineev}(2006)}]{PhysRevB.74.144522}%
  \BibitemOpen
  \bibfield  {author} {\bibinfo {author} {\bibfnamefont {M.}~\bibnamefont
  {Houzet}}\ and\ \bibinfo {author} {\bibfnamefont {V.~P.}\ \bibnamefont
  {Mineev}},\ }\href {\doibase 10.1103/PhysRevB.74.144522} {\bibfield
  {journal} {\bibinfo  {journal} {Phys. Rev. B}\ }\textbf {\bibinfo {volume}
  {74}},\ \bibinfo {pages} {144522} (\bibinfo {year} {2006})}\BibitemShut
  {NoStop}%
\bibitem [{\citenamefont {Asano}\ and\ \citenamefont
  {Golubov}(2018)}]{asano:prb2018}%
  \BibitemOpen
  \bibfield  {author} {\bibinfo {author} {\bibfnamefont {Y.}~\bibnamefont
  {Asano}}\ and\ \bibinfo {author} {\bibfnamefont {A.~A.}\ \bibnamefont
  {Golubov}},\ }\href {\doibase 10.1103/PhysRevB.97.214508} {\bibfield
  {journal} {\bibinfo  {journal} {Phys. Rev. B}\ }\textbf {\bibinfo {volume}
  {97}},\ \bibinfo {pages} {214508} (\bibinfo {year} {2018})}\BibitemShut
  {NoStop}%
\bibitem [{\citenamefont {Sato}\ and\ \citenamefont
  {Asano}(2020)}]{takumi:prb2020}%
  \BibitemOpen
  \bibfield  {author} {\bibinfo {author} {\bibfnamefont {T.}~\bibnamefont
  {Sato}}\ and\ \bibinfo {author} {\bibfnamefont {Y.}~\bibnamefont {Asano}},\
  }\href {\doibase 10.1103/PhysRevB.102.024516} {\bibfield  {journal} {\bibinfo
   {journal} {Phys. Rev. B}\ }\textbf {\bibinfo {volume} {102}},\ \bibinfo
  {pages} {024516} (\bibinfo {year} {2020})}\BibitemShut {NoStop}%
\bibitem [{Note3()}]{Note3}%
  \BibitemOpen
  \bibinfo {note} {Changing $V_0$ in the horizontal axis is realized by
  applying an external Zeeman field in addition to the magnetic moment
  possessed in a ferromagnet.}\BibitemShut {Stop}%
\bibitem [{\citenamefont {Oreg}\ \emph {et~al.}(2010)\citenamefont {Oreg},
  \citenamefont {Refael},\ and\ \citenamefont {von Oppen}}]{oreg:prl2010}%
  \BibitemOpen
  \bibfield  {author} {\bibinfo {author} {\bibfnamefont {Y.}~\bibnamefont
  {Oreg}}, \bibinfo {author} {\bibfnamefont {G.}~\bibnamefont {Refael}}, \ and\
  \bibinfo {author} {\bibfnamefont {F.}~\bibnamefont {von Oppen}},\ }\href
  {\doibase 10.1103/PhysRevLett.105.177002} {\bibfield  {journal} {\bibinfo
  {journal} {Phys. Rev. Lett.}\ }\textbf {\bibinfo {volume} {105}},\ \bibinfo
  {pages} {177002} (\bibinfo {year} {2010})}\BibitemShut {NoStop}%
\bibitem [{\citenamefont {Lutchyn}\ \emph {et~al.}(2010)\citenamefont
  {Lutchyn}, \citenamefont {Sau},\ and\ \citenamefont
  {Das~Sarma}}]{lutchyn:prl2010}%
  \BibitemOpen
  \bibfield  {author} {\bibinfo {author} {\bibfnamefont {R.~M.}\ \bibnamefont
  {Lutchyn}}, \bibinfo {author} {\bibfnamefont {J.~D.}\ \bibnamefont {Sau}}, \
  and\ \bibinfo {author} {\bibfnamefont {S.}~\bibnamefont {Das~Sarma}},\ }\href
  {\doibase 10.1103/PhysRevLett.105.077001} {\bibfield  {journal} {\bibinfo
  {journal} {Phys. Rev. Lett.}\ }\textbf {\bibinfo {volume} {105}},\ \bibinfo
  {pages} {077001} (\bibinfo {year} {2010})}\BibitemShut {NoStop}%
\end{thebibliography}%

\end{document}